%% file: main_IEEE.tex
\documentclass[conference]{IEEEtran} 
\usepackage[T2A,T1]{fontenc}
\usepackage[utf8]{inputenc}
\usepackage[dvipsnames]{xcolor}
\usepackage{textgreek} 
\usepackage{algorithm}
\usepackage[noend]{algpseudocode} 

\usepackage{rotating} 
\usepackage{graphicx} 

\usepackage[colorlinks=true,linkcolor=blue,breaklinks=True,citecolor=brown,urlcolor=blue]{hyperref}

\usepackage{dingbat}

\usepackage{adjustbox}
\usepackage[linesnumbered,algo2e,ruled]{algorithm2e}
\usepackage{multirow}
\usepackage{booktabs,subcaption,dcolumn}
\usepackage{tikz}
\usepackage{enumitem}
\setlist[itemize]{noitemsep, topsep=0pt}
\usepackage[normalem]{ulem}
\usepackage{tabularx}
\newcolumntype{Y}{>{\raggedright\arraybackslash}X} 
\usepackage{setspace}

\usepackage{pifont} 
\usepackage{svg}
\usepackage{soul}
\usepackage[font=footnotesize]{caption}
\usepackage[noadjust]{cite}

\usepackage{amsmath}
\usepackage{amsfonts, amssymb}

\usepackage{colortbl}

\usepackage{tablefootnote}
\usepackage[most]{tcolorbox}
\usepackage[sort&compress, numbers]{natbib}
\usepackage{booktabs}
\usepackage{siunitx}

\AtBeginDocument{%
  \providecommand\BibTeX{{%
    \normalfont B\kern-0.5em{\scshape i\kern-0.25em b}\kern-0.8em\TeX}}}

\usepackage{listings}
\input{commands.tex}

\begin{document}

\input{title}

\input{authors.tex}

\pagestyle{plain}
\maketitle

\input{sections/0-abstract}

\input{structure}

\bibliographystyle{IEEEtran}

{\footnotesize
\input{main.bbl}

}

 \appendices
\input{appendix/structure_app}

\end{document}

%% file: commands.tex
\newcolumntype{?}{!{\vrule width 1.5pt}}

\newcommand{\cyrillic}[1]{{\begingroup\fontencoding{T2A}\selectfont #1\endgroup}}

\newtcolorbox{cooltextbox}[1][]{%
    colback=black!5,
    colframe=black!5,
    notitle,
    sharp corners,
    borderline west={0pt}{0pt}{red!80!black},
    enhanced,
    breakable,
    left=0pt,
    right=0pt,
    top=0pt,
    bottom=0pt
    }

\newcommand\smamath[1]{{\small $#1$}}

\newcommand\revision[1]{%
  \bgroup
  \hskip0pt\color{blue!80!black}%
  #1%
  \egroup
}

%% file: title.tex
\title{Adversarial News and Lost Profits: Manipulating Headlines in LLM-Driven Algorithmic Trading}

%% file: authors.tex
\author{
\hspace{-0cm}
\IEEEauthorblockN{Advije Rizvani\IEEEauthorrefmark{2}, Giovanni Apruzzese\IEEEauthorrefmark{2}\IEEEauthorrefmark{5}, Pavel Laskov\IEEEauthorrefmark{2}
\\}
\IEEEauthorblockA{{ 
\IEEEauthorrefmark{2}\textit{Liechtenstein Business School -- University of Liechtenstein},
\IEEEauthorrefmark{5}\textit{Dept. of Computer Science -- Reykjavik University},
}}

{\small 
\{name.surname\}@uni.li 
}
}

%% file: sections/0-abstract.tex
\begin{abstract}
Large Language Models (LLMs) are increasingly adopted in the financial domain. Their exceptional capabilities to analyse textual data make them well-suited for inferring the sentiment of finance-related news. Such feedback can be leveraged by algorithmic trading systems (ATS) to guide buy/sell decisions. However, this practice bears the risk that a threat actor may craft ``adversarial news'' intended to mislead an LLM. In particular, the news headline may include ``malicious'' content that remains invisible to human readers but which is still ingested by the LLM. Although prior work has studied textual adversarial examples, their system-wide impact on LLM-supported ATS has not yet been quantified in terms of monetary risk. 

To address this threat, we consider an adversary with no direct access to an ATS but able to alter stock-related news headlines on a single day. We evaluate two human-imperceptible manipulations in a financial context: Unicode homoglyph substitutions that misroute models during stock-name recognition, and hidden-text clauses that alter the sentiment of the news headline. We implement a realistic ATS in Backtrader that fuses an LSTM-based price forecast with LLM-derived sentiment (FinBERT, FinGPT, FinLLaMA, and six general-purpose LLMs), and quantify \textit{monetary impact} using portfolio metrics. Experiments on real-world data show that manipulating a one-day attack over 14 months can reliably mislead LLMs and reduce annual returns by up to 17.7 percentage points. To assess real-world feasibility, we analyze popular scraping libraries and trading platforms and survey 27 FinTech practitioners, confirming our hypotheses. We notified trading platform owners of this security issue.
\end{abstract}

%% file: structure.tex
\input{sections/1-introduction}
\input{sections/2-background}

\input{sections/3-threat}

\input{sections/4-ATS}

\input{sections/5-evaluation}

\input{sections/6-cross-model}

\input{sections/7-survey}

\input{sections/8-discussion}
\input{sections/9-mitigations}

\input{sections/10-conclusions}

\input{sections/LLMusage}

%% file: sections/1-introduction.tex
\section{Introduction}
\label{sec:introduction}

\noindent
Financial markets are moved by information. 
In such a highly-competitive ecosystem, each player tries to be the first to make a good deal. Therefore, being quick at obtaining, processing, and using information from the most recent news is crucial to keep generating a profit~\cite{schumaker2009quantitative,mitra2011applications, fedyk2024front}.
Over the past decade this process has become increasingly automated~\cite{alzheev2020comparative,AlgorithmicTrading,milionis2024automated_fc}. Modern algorithmic trading systems (ATS) continuously ingest news streams from vendors (e.g., Refinitiv, Bloomberg~\cite{bloomberggpt,refinitiv-marketpsych}), or public sources~\cite{akhila2025hybrid, bhat2024stock}. The headlines of such news can be mapped to specific stocks, analyzed using machine-learning (ML) models such as FinBERT~\cite{huang2023finbert}, FinGPT~\cite{yang2023fingptopensourcefinanciallarge}, FinLLaMa~\cite{iacovides2024finllama}, or ChatGPT, and integrated into decision-making pipelines~\cite{iacovides2025findpo, frattini2022financial, yadava2024impact, akhila2025hybrid}.  
If sentiment extraction is accurate, ATS enable their users to exploit new information and gain profit.

Existing news-driven ATS pipelines implicitly assume that the textual input received by the large-language model (LLM) is trustworthy. This assumption is fragile. First, even the established sources of financial data can be manipulated, causing billion dollar losses, as in the case of the AP Twitter hack \cite{karppi2016social}, or facilitating insider trading and securities fraud, exemplified by the Emulex incident \cite{FowlerFranklinHyde2001}. As recently hypothesized by Boucher et al.~\cite{boucher2022bad}, ``a dishonest company could mask negative information in its financial filings so that the specialist search engines used by stock analysts fail to pick it up.'' Put simply, there is plenty of evidence showing that financial news cannot be trusted. Hence, we wonder (RQ): \textit{what happens to the profitability of an ATS that has ingested an ``adversarial news'', deliberately manipulated so that it misleads an LLM while remaining visually unaltered to the human eye?}

We carried out a systematic literature review encompassing over 25k papers (§\ref{ssec:ml-finance-gap}). Despite many works showing that human-imperceptible textual changes can mislead various NLP-based methods~\cite{boucher2022bad}, including LLMs~\cite{kulkarni2025ml}; and while prior work has investigated the impact that adversarial perturbations may have on ATS driven by stock-price forecasting models (e.g., LSTM~\cite{rizvani2025ephemeral}), we could not find any work that evaluated the effects that such ``adversarial news'' may have on a full-fledged LLM-driven ATS. If the LLM fails, how much \$ does the ATS lose? Answering such a question is beneficial for researchers and professionals alike: depending on the economical losses, one can determine whether it is sensible to develop, deploy, and maintain, specific defenses.

\begin{figure*}[t]
  \centering
  \subfloat[Unicode homoglyph substitution.\label{fig:tA}]{
    \includegraphics[width=.485\textwidth]{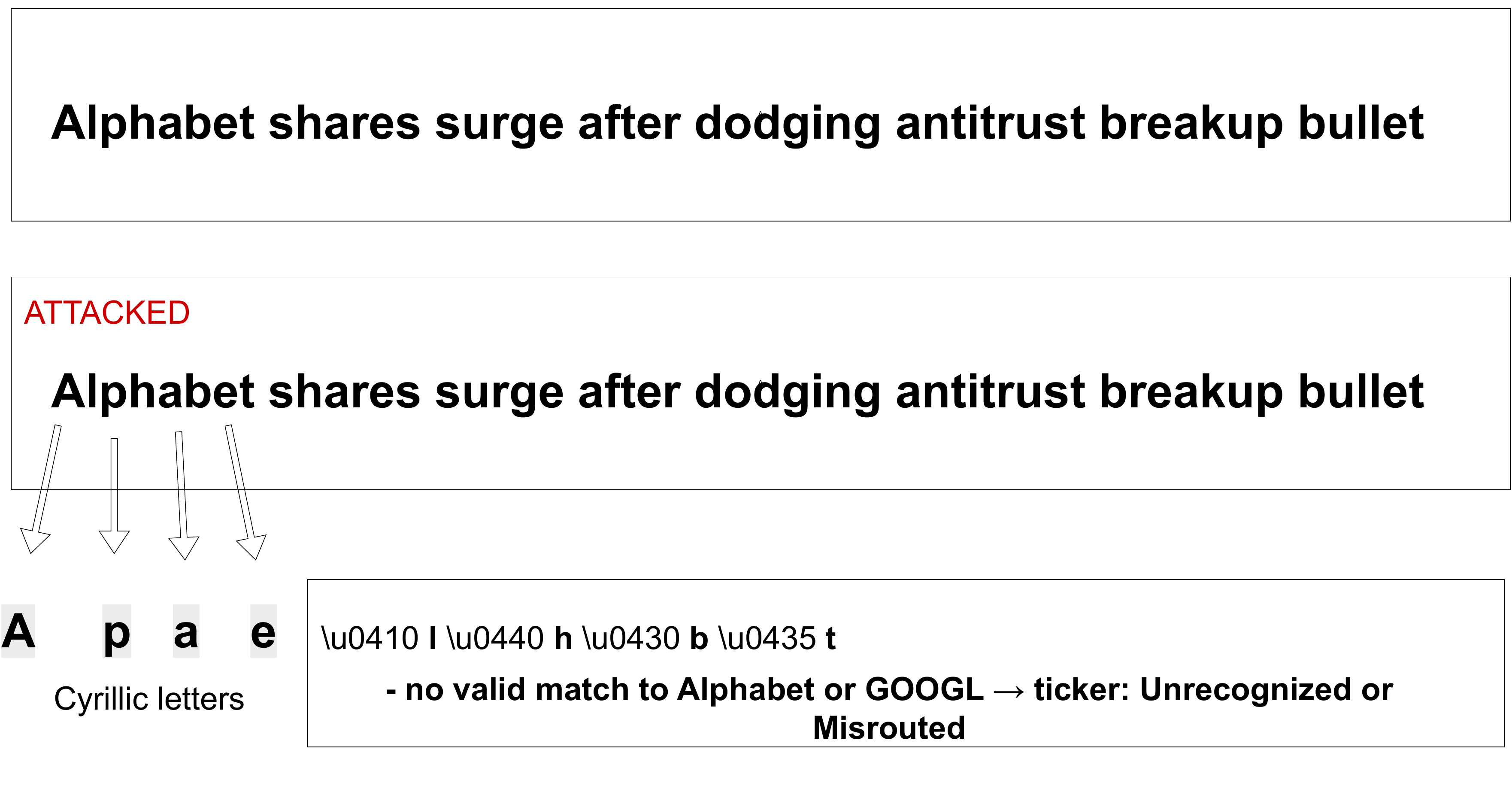}}
  \hfill
  \subfloat[Hidden-text injection.\label{fig:tB}]{
    \includegraphics[width=.485\textwidth]{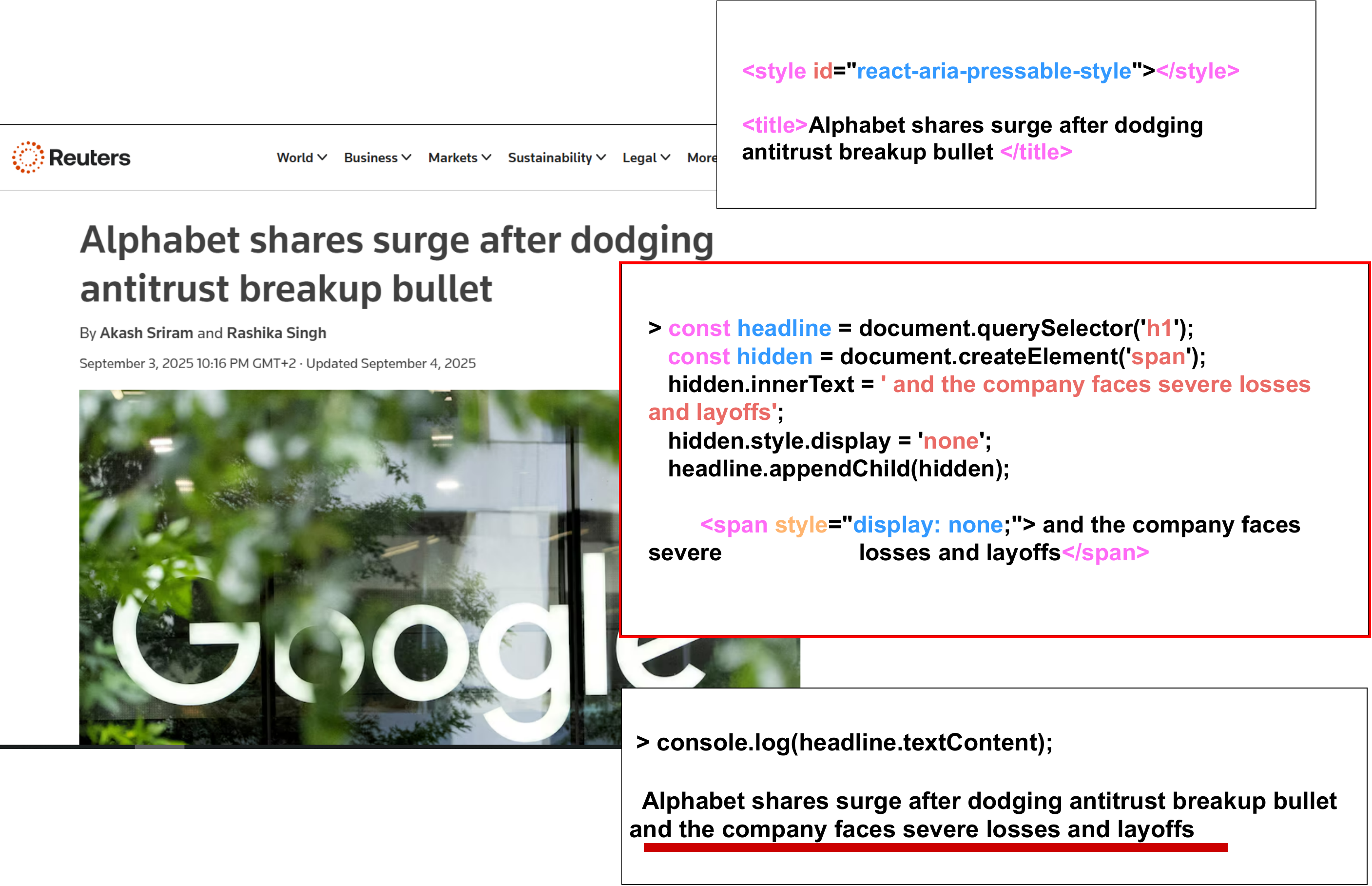}}
  \caption{\textbf{Two stealth edits that humans do not notice, but trading models do.}
  (a) Mixed-script Unicode breaks stock mapping. (b) A hidden clause with
  \texttt{display:none} is invisible to users but parsed by the model. (the original news article can be found at:~\cite{reuters2025alphabet}) 
  }
  \label{fig:teaser}
  \vspace{-5mm}
\end{figure*}

There are various ways to leverage imperceptible textual manipulations~\cite{boucher2022bad,boucher2025vision,hackett2025bypassing} and apply them to financial news. To provide an exemplary answer to our RQ, we consider two complementary techniques to deceive LLM-driven ATS, both of which focus on tampering with the headline. Specifically:
\begin{itemize}[leftmargin=*,noitemsep,topsep=0pt]
  \item \textit{Unicode homoglyph substitution.} A few characters in a stock name (e.g., \texttt{A}, \texttt{e}) can be replaced by visually indistinguishable Unicode counterparts (e.g., Cyrillic ``{\begingroup\fontencoding{T2A}\selectfont A\endgroup}'' \cyrillic{А}, ``{\begingroup\fontencoding{T2A}\selectfont е\endgroup}'' \cyrillic{е}).  
  The manipulated headline appears to be unchanged to the human eye. When such headline is used in the ATS pipeline, it may elicit wrong decisions in the stock mapping algorithm, determining which stock a given headline refers to, and thus ``misroute'' a headline. This attack is illustrated in Fig.~\ref{fig:teaser}(a).

  \item \textit{Hidden-text injection.} Additional text, which reverses the perceived sentiment, can be inserted into the headline and wrapped in {\footnotesize \texttt{<span style="display:none">... </span>}}.
  Obviously, a human would only see the original headline, whereas the model ingests the hidden content and predicts the wrong sentiment. This attack is illustrated in Fig.~\ref{fig:teaser}(b).
\end{itemize}
\noindent
Both of the aforementioned techniques can be reliably applied from outside the ATS, e.g., by a rogue editor (§\ref{sec:threat}). 

We measure the impact of such adversarial news on a realistic ATS. We take the open-source ATS proposed in a recent work~\cite{rizvani2025ephemeral} and enhance it by integrating an LLM-driven module that processes news headlines (inspired by~\cite{akhila2025hybrid}). We empirically verify, via a 14-months-long simulation, that our ATS outperforms the one in~\cite{rizvani2025ephemeral}. Then, for our RQ, we measure the loss, in terms of relative drop in cumulative returns, incurred by our ATS when it ingests \textit{adversarial news on a single day}. On average, the drop is \smamath{\approx}3.5\%. However, the ATS keeps generating a profit, meaning that the targeted organization is likely oblivious of having incurred such a loss.

\vspace{2mm}
\noindent
This work hence makes the following contributions: 

\begin{itemize}[leftmargin=*,noitemsep,topsep=0pt]
    
    \item \textbf{System-wide attack evaluation.} First, we implement a custom ATS (§\ref{sec:ats}) that combines a news-driven LLM for sentiment analysis of headlines with an LSTM for stock-price forecasting. We show that our ATS outperforms prior work~\cite{rizvani2025ephemeral}, justifying the validity of our considered ATS. Then, we assess such an ATS against ``adversarial news'' entailing human-imperceptible manipulations of headlines, exploiting homoglyph substitutions and invisible text.

    \item \textbf{Quantification of Economic Impact.} Our assessment (§\ref{sec:evaluation}) reveals that our considered LLM (FinBERT) fails to correctly analyse our adversarial news 99\% of the time for the homoglyph attack, and 67\% of the time for the invisible HTML text. Then, we measure the financial loss that such attacks may have on the ATS when it is subject to adversarial news for just one day across a 14-month simulation. We find that the profitability decreases by \smamath{\approx}3\%. Transferability experiments (§\ref{sec:cross}) on 9 other LLMs (e.g., FinGPT, O3) show that the attack is effective even against similar ATS.

 \item \textbf{Real-world Validation and Recommendations.} Through a user study (n=27) with FinTech practitioners (§\ref{sec:survey}), we provide evidence that our design choices align with the real-world, and that our threat model is a realistic risk~(§\ref{ssec:applicability}). Finally, we elaborate on possible countermeasures (§\ref{sec:mitigations}).
\end{itemize}
We also systematically analyse over 25k papers from top-tier ML/NLP and Security venues, proving that we are the first to tackle this problem (§\ref{sec:background}). We share our code repository~\cite{ourRepo}.

%% file: sections/2-background.tex
 \section{Background and Related work}
\label{sec:background}
\noindent
The core problem addressed in our work lies at the confluence of finance, machine learning (ML) and security. To articulate the scientific underpinnings of our contribution, we review the role of ML and LLMs in finance~(\S\ref{ssec:ml-llm-finance}), establish a connection of our work to prior results in adversarial ML~(\S\ref{ssec:ml-security}), and elucidate, via a systematic literature review, the current gap in applying ML security analysis in the financial domain~(\S\ref{ssec:ml-finance-gap}).

\subsection{ML and LLM in Finance}
\label{ssec:ml-llm-finance}

\noindent
Since 1990s, ML has gradually transformed the practice of quantitative finance.  
Classical statistical methods such as ARIMA have long been used for time-series modeling in forecasting stock prices, volatility, and macroeconomic indicators~\cite{ariyo2014stock, tang2022way, peter2012arima}.  
However, these methods rely on stationarity assumptions and limited memory, restricting their ability to capture nonlinear dynamics, as well as long(er)-term trends. 

The advances in ML and especially deep learning (DL) introduced models better suited to sequential data.
Recent studies show that ML models, in particular, LSTMs 
\emph{outperform classical time-series baselines} (e.g., ARIMA/VAR) on equity prediction under rolling walk-forward evaluation by better capturing temporal dependencies \cite{selvin2017stock,ding2020study,sen2022long}. 
More recently, Transformer-based architectures have been shown to outperform non-Transformer models, including LSTMs, in financial time series forecasting~\cite{zhang2022transformer,li2025enhancing}. 
For example, Yanez et al.~\cite{yanez2024stock} benchmarked LSTM and Transformer models on ten real-world financial time-series datasets (daily stock indices). 
Transformers consistently outperformed LSTMs, achieving lower prediction error on 8 out of 10 datasets in terms of RMSE and on 9 out of 10 in terms of MAE. 
As an illustration, on one dataset the RMSE dropped from 0.0142 (LSTM) to 0.0118 (Transformer), and the MAE from 0.0099 to 0.0085, corresponding to relative improvements of about 15--17\%.

Besides forecasting models driven by numerical data (e.g., stock prices), the availability of vast unstructured text data such as company reports, or news headlines, has inspired the deployment of natural language processing (NLP) in finance. 
Finance-tuned language models (e.g., FinBERT, FinGPT, FinLlama) are trained on finance domain to better capture task-specific semantics ~\cite{choi2025large,iacovides2024finllama}. 
On benchmark sentiment datasets, such as Financial PhraseBank and FiQA, domain-specific LLMs consistently outperform general models: on the former, FinMA-7B achieves 88\% accuracy, compared to 78\% for GPT-4; on the latter, the best finance-specific model reaches the F1 score of 0.79, closely matching larger general-purpose models~\cite{xie2024finben}. 
Proprietary models such as BloombergGPT~\cite{bloomberggpt} further indicate industry commitment (and investment) in adapting large-scale architectures to financial tasks~\cite{wu2023bloomberggpt}.

\subsection{Security of ML and LLM}
\label{ssec:ml-security}

\noindent
Given the risk of immense losses, showcased by the past data manipulation incidents in finance~\cite{karppi2016social}, a question arises: what if similar disasters happen to ML-driven methods for computational finance? The assessment of such risks requires a connection to general principles of ML security, which have been intensely explored since the discovery of the so-called adversarial perturbations~\cite{biggio2012poisoning,biggio2013evasion,szegedy2014intriguing}. 

The crucial insight emerging from ML security literature is the importance of realistic threat modeling \cite{vsrndic2014practical,biggio2018wild}. The majority of academic attacks and defenses revolve around the access to the gradient of the learned model \cite{athalye2018obfuscated}. In practice, attackers often employ simpler, cost-driven strategies~\cite{apruzzese2023real}.  Furthermore, quantification of risks arising from adversarial perturbations should not be limited to ML-intrinsics cost functions such as accuracy, AUC or F1-score; in many applications domains, different, even physical cost functions are of greater importance, see, e.g., a discussion in \cite{apruzzese2022wild}. Prior systems work also shows that ML vulnerabilities may lead not only to higher costs, but to operational failures as well~\cite{grosse2017adversarial,shumailov2021sponge, demetrio2021functionality}.  

Recent developments in LLM reveal additional attack surfaces.  
Prompt injection and jailbreak techniques demonstrate how models can be steered into ignoring intended instructions or producing unsafe outputs~\cite{greshake2023not,zou2023universal}.
Markup-based manipulations, such as injecting hidden HTML or CSS tokens, exploit the mismatch between human-visible text and model-parsed content~\cite{brach2025ghosts}. 
Unicode homoglyph substitutions further show how visually indistinguishable characters can alter entity recognition or routing in downstream systems~\cite{pajola2021fall,sarabamoun2025special}.  

Another tenet of ML/LLM security is the necessity to trace how model errors propagate through end-to-end systems and their decision logic \cite{arp2022and}. System-level analyses have shown that ``robustness to adversarial noise'' that appears to be strong in ``laboratory conditions'' can degrade in realistic deployment conditions with temporal non-stationarity, and small model errors can cascade into operational failures \cite{pendlebury2019tesseract}. In finance, this translates directly into monetary impact; hence, adversarial effects must be necessarily studied at the system level~\cite{rizvani2025ephemeral}.

\subsection{Research Gap (and Systematic Literature Review)}
\label{ssec:ml-finance-gap}    

\noindent
To date, the applications of ML security in finance remain underexplored. The majority of prior works target price time-series perturbations or forecasting dataset poisoning~\cite{rizvani2025ephemeral,chen2021adversarial,gallagher2022investigating}. However, implications of these findings for LLM-driven ATS remain unclear. Specifically, there is lack of threat models and \emph{system-level} evaluations of adversarial manipulations targeting financial LLM news ingestion, as well as measurements of their downstream economic impact.

A systematic review in~\cite{rizvani2025ephemeral} across 10 years (2013--2023), identified that, among \smamath{\approx}7k papers, only 6~\cite{chen2021adversarial,dang2020adversarial,mode2020adversarial,goldblum2021adversarial,nehemya2021taking,gallagher2022investigating} considered ATS under adversarial manipulation. Yet, none of these 6 papers consider ATS whose decision making leverages LLMs analysing financial news. So, to confirm if, as of September 2025, the security of LLMs for news ingestion is still an underexplored research area, we carried out another -- larger -- systematic literature review. First, and similarly to~\cite{rizvani2025ephemeral}, we examined {2,169} papers from top-tier security venues (e.g., USENIX Security, IEEE S\&P, CCS, NDSS); within 2024--2025, because the previous years had been covered in~\cite{rizvani2025ephemeral}); then, we analyse an additional set of 23,038 papers published within 2020--2025 in top-tier ML venues (NeurIPS, ICML, ACL, EMNLP); we further complement with snowballing~\cite{wohlin2014guidelines} and a broader search on Google Scholar. The entire systematic procedure is reported in the Appendix~\ref{app:slr} (since ours is not a review paper, we deferred this content to the Appendix). Despite our extensive search, we were unable to find any work on the security of ML in ATS beyond those found in~\cite{rizvani2025ephemeral}.  

Finally, while new financial LLM benchmarks have appeared~\cite{xie2024finben}, no adversarial ML attack was tested against them. However, as hypothesized in~\cite{boucher2022bad}, as well as in the (unpublished) work by similar authors~\cite{deza2020robustness} which did not consider LLMs, we have reason to believe that LLMs for news ingestion can be deceived via human-imperceptible manipulations.
These findings motivate us to explore this research gap.

%% file: sections/3-threat.tex
\section{Threat Model and Proposed Attack}
\label{sec:threat}

\noindent
As the starting point of our contribution, we formalize our proposed threat model. We first outline the target system~(§\ref{ssec:target-system}); then describe the envisioned attacker~(§\ref{ssec:attacker}) and finally compare our threat model to those of prior related work~(§\ref{ssec:prior}).

\begin{figure}[t]
  \centering
  \includegraphics[width=\columnwidth]{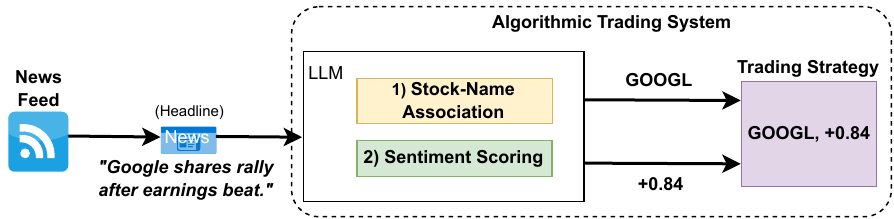} 
 \caption{\textbf{Extraction of stock-name association and sentiment from a headline.} The ATS receives, as input, the headline of a given news, which is then processed via LLM(s), and used to make trading decisions.}

  \label{fig:news-two-step}
  \vspace{-5mm}
\end{figure}

\noindent
\subsection{Target System}
\label{ssec:target-system}

\noindent
The targeted system is an ATS that follows a given portfolio of stocks (e.g., GOOGL, AMZN) according to any given trading strategy, and that is assumed to yield a profit to its owners. 

The trading decisions of the ATS are driven by two signals:
{\small \textit{(i)}}~a \emph{price} forecast from a univariate model (e.g., LSTM~\cite{rizvani2025ephemeral}) over daily bars, and  {\small \textit{(ii)}}~a \emph{text} signal obtained by mapping headlines of financial news to a per-stock sentiment score. 

Such headlines, which can be received either by vendors (e.g.,~\cite{refinitiv-eikon}) or scraped autonomously by the ATS owners (e.g., the work in~\cite{deza2020robustness} considered tweets on Twitter), are assumed to be processed in two steps.
First, a \emph{stock-name association} module assigns each headline to the most fitting 
company (so-called ``ticker'') in the portfolio. 
Second, a \emph{sentiment scoring} mechanism processes the headline to produce a polarity \(s_{t,\text{stock}}\in[-1,1]\) for day \(t\). Both of these operations are carried out by an LLM (which have been shown to excel at similar tasks~\cite{iacovides2024finllama}). See Fig.~\ref{fig:news-two-step} for a schematic of how the ATS elaborates a given headline. The ATS records \((\text{stock name}, s_{t,\cdot})\) and later fuses the (smoothed) sentiment with the price signal (e.g.,~\cite{akhila2025hybrid}) to issue buy/hold/sell actions.

\begin{figure}[t]
  \centering
  \includegraphics[width=0.9\columnwidth]{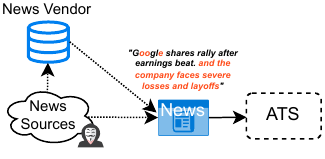}
  \vspace{-4mm}
  \caption{\textbf{Threat Model.} The attacker, outside the ATS, manipulates the headline of a single news mentioning a stock within the portfolio of the targeted ATS. Text in red denotes the adversarial manipulation: ``Google'' is written in homoglyphs; or an (invisible) sentence is added to force a negative sentiment. (Note: in this work, we consider either the homoglyph or the invisible text, not both---though a real attacker can certainly combine both methods.)}
  \label{fig:threat}
  \vspace{-5mm}
\end{figure}

\subsection{The Attacker}
\label{ssec:attacker}

\noindent
We describe our attacker according to the recommendations in~\cite{apruzzese2023real}, which endorse to adopt a system-wide view:

\begin{itemize}[leftmargin=*,noitemsep,topsep=0pt]
    \item \textit{Goal:} The attacker seeks to induce the targeted ATS to yield an inferior revenue to its owners (as measured by a reduced cumulative returns of the ATS over a given time period). 

    \item \textit{Knowledge.} The adversary knows that the ATS ingests headlines of financial news, links them to tickers, and derives a sentiment score (e.g., this is a reasonable assumption since it is a common practice~\cite{bhat2024stock, akhila2025hybrid}). The adversary also knows at least one stock $s^\star$ traded by the ATS (e.g., GOOGL), inferred from, e.g., prior trades. The attacker does not know low-level details about the internal components of the ATS (e.g., parameters/training data of the ML models, full portfolio, or price feeds): such information is confidential~\cite{rizvani2025ephemeral}.

    \item \textit{Capabilities:} The attacker can manipulate news headlines mentioning $s^\star$. This can occur either: {\small \textit{(a)}}~by deliberately writing an ``adversarial'' news at the source---as hypothesized in~\cite{boucher2022bad}; or {\small \textit{(b)}}~by a compromise of the news publishing  pipeline---e.g., via a man-in-the-middle attack between the source and the ATS~\cite{nehemya2021taking,rizvani2025ephemeral}); though very unlikely, it can also happen {\small \textit{(c)}}~at the vendor---under the assumption that the ATS uses vendor-provided news, instead of scraping them from public sources. Changes must be imperceptible to humans. 
    To ensure stealthiness, the attacker can attempt such a manipulation \textit{only on a specific day} $t^\star$.
    The attacker cannot touch prices, retrain models, manipulate the news beyond the headline, or persist across multiple days. 
\end{itemize}
\noindent
We examine two practical \textit{strategies} within this threat model:

\begin{itemize}[leftmargin=*,noitemsep,topsep=0pt]
    \item \textbf{Unicode homoglyph substitution:} The attacker replaces visually identical Unicode characters (e.g., Latin ``A'' with Cyrillic {\begingroup\fontencoding{T2A}\selectfont \endgroup}``\cyrillic{А}'') in the company name, potentially inducing the stock-name association LLM to fail to assign a stock to such a headline (leading to a missed opportunity).

   \item \textbf{Hidden-text manipulation:} The attacker adds sentiment-reversing phrases using invisible HTML tags (e.g., {\scriptsize \texttt{<span style="display:none">and the company faces severe losses and layoffs </span>}}, which can also be concealed by setting {\scriptsize \texttt{style="font-size: 0pt"}}), inducing the sentiment-scoring LLM to assign the wrong polarity to an otherwise positive headline, which may lead the ATS to incorrectly sell/hold a stock that would otherwise be bought.

\end{itemize}
Practically implementing such ``imperceptible'' changes is trivial for an attacker that can tamper with the headlines. We visualize our envisioned scenario in Fig.~\ref{fig:threat}. We provide real-world considerations on our threat model in §\ref{ssec:applicability}.

\subsection{Comparison to Prior Threat Models (Novelty)}
\label{ssec:prior}

\noindent
Most adversarial ML research in finance envisions attackers with white-box access, control over training data, or influence across long historical horizons~\cite{liu2018data,chen2021adversarial,nasr2019comprehensive,gallagher2022investigating,goldblum2021adversarial}. While these assumptions help benchmark model robustness, they are hard to meet in the real world: ATS are highly-secure systems~\cite{rizvani2025ephemeral}. 

Prior work~\cite{rizvani2025ephemeral} has introduced realistic, time-constrained adversarial perturbations targeting ATS exclusively reliant on price inputs (processed via LSTM). 
In contrast, we consider ATS that combine numerical price inputs with textual news ingestion, and we specifically target LLM (i.e., NLP-based models) instead of LSTM models for stock-price forecasting.

While adversarial threats to LLMs (such as prompt injection~\cite{greshake2023not} or Unicode-based misdirection~\cite{pajola2021fall,cooper2025lies}), have been studied in the generic NLP context, we are not aware of any work scrutinizing the system-wide impact of these threats to an ATS, including for the specific purpose of sentiment-driven decision-making of financial trades. Moreover, even though early (unpublished) works considered attacks against NLP-based techniques~\cite{deza2020robustness} for financial applications, while others hypothesized that news could be adversarially manipulated to affect financial systems~\cite{boucher2022bad}, the economical impact that similar attacks can have on LLM-driven ATS has never been examined. Such a gap motivates our system-wide evaluation.

%% file: sections/4-ATS.tex
\section {System Implementation}
\label{sec:ats}

\noindent
To assess the impact of our threat model, we implement an ATS resembling the targeted system. We  first present the overall design of the ATS (§\ref{ssec:design}), then describe the development of the underlying ML models (§\ref{ssec:models}), and conclude by measuring the baseline performance of our ATS (§\ref{ssec:baseline}).

\subsection{Design of our Algorithmic Trading System (ATS)}
\label{ssec:design}
\noindent
ATS are systems\footnote{\textbf{Primer on ATS.} In principle, ATS receive some signals that are used to carry out trading decisions (e.g., buy or sell a stock), provided that enough resources are available. Such decisions are dictated by the specified trading strategy and overarching portfolio. The performance of an ATS can be computed by measuring the cumulative returns over a certain time period.} that automate the ``predict–decide–trade'' loop~\cite{nuti2011algorithmic}, ideally yielding a profit to their owners.

\begin{figure}[t]
  \centering
  \includegraphics[width=\columnwidth]{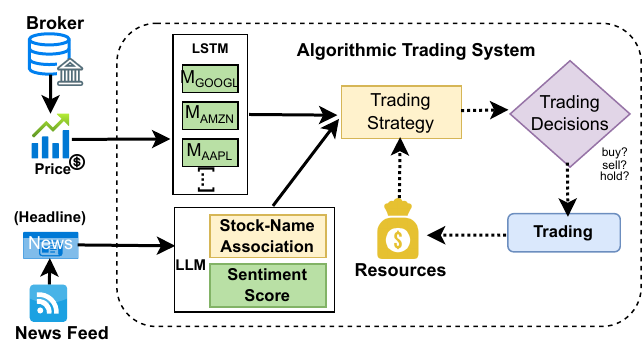}
  \vspace{-6mm}
  \caption{\textbf{Schema of an ATS with parallel price and news streams.} 
  Market data and financial headlines are processed by deep learning models 
  (LSTM for prices, LLM for news). Their outputs feed a decision module that issues daily buy/hold/sell actions 
  under resource constraints.}
  \label{fig:ats}
  \vspace{-4mm}
\end{figure}

We provide a schema of our envisioned ATS in Fig.~\ref{fig:ats}.
 Recall that, in our setting, the ATS integrates two heterogeneous signals for each stock in the portfolio: 
{\small \textit{(i)}}~a price forecast from LSTM models over historical bars, and 
{\small \textit{(ii)}}~a sentiment score extracted by LLMs from news headlines. Our ATS obtains the input data to generate such signals by the following sources:

\begin{itemize}[leftmargin=*,noitemsep,topsep=0pt]
  \item \textit{Stock market data (analysed by the LSTM):} we use daily OHLCV records (open, high, low, close, and traded volume) provided by YahooFinance~\cite{yahooFinance} during 2013--2025.

  \item \textit{News data (analysed by the LLM).} Headlines are drawn from Refinitiv~\cite{refinitiv-eikon}, a well-known vendor of financially-related news. Note, however, that our threat model (§\ref{ssec:target-system}) does not strictly assume the presence of a news vendor. Indeed, we use Refinitiv because it provides (under a payment) a curated database of historical news that are relevant for our experiments. In practice, the exact same data could be obtained, e.g., by monitoring/subscribing to financial feeds (e.g., Reuters), or by scraping the Web.\footnote{For instance, by querying the \textit{get\_news\_headlines} API (documented in~\cite{refinitiv2020api}) for headlines referring to news about NVDA, one result we get is ``Nvidia shares surge 13\%, lift market value a record \$330 billion'', which is the exact same headline reported by the news source (available at~\cite{reuters2024nvidia})}
\end{itemize}
(Unfortunately, we cannot release data from Refinitiv, but the headlines always refer to publicly-available news.)

The overarching design of our ATS is rooted on prior peer-reviewed work (we are not aware of open-source ATS available for security research and used by real-world companies). At most one decision per asset per day is made~\cite{rizvani2025ephemeral}. 
Signals observed at the end of day $t$ translate into orders that execute at the market open of day $t{+}1$, enforcing a clear temporal boundary and preventing look-ahead bias~\cite{baron2023measuring,wang2022information}.
Trading is subject to transaction costs and capital constraints, applied across all runs~\cite{milionis2024automated_fc}.  
According to~\cite{rizvani2025ephemeral} (also validated with a user study), such an ATS resembles a realistic setup. Finally, combining LSTM with transformer-based models (e.g., FinBERT) for sentiment analysis was also proposed in prior work~\cite{akhila2025hybrid} (unfortunately, the code of~\cite{akhila2025hybrid} is not public, which is why we have to develop this component from scratch). 

To further align our ATS to real-world deployments, our trading strategy is implemented in Backtrader, a professional backtesting engine~\cite{backtrader} (not used in~\cite{rizvani2025ephemeral}), and is configured to trade a diversified portfolio of ten large-cap U.S. equities: \{{\small GOOGL, AAPL, NVDA, MSFT, AMZN, META, TSLA, LLY, JPM, XOM}\}.
These assets were selected for their liquidity, sectoral diversity, and high frequency of news coverage~\cite{tradingviewLargeCap}. Put simply, our ATS not only follows our envisioned ``target system'' (refer to §\ref{ssec:target-system}) but also resembles a realistic setup, thereby enabling us to carry out a meaningful assessment of attacks stemming from our threat model.

\subsection{Developing and Orchestrating the ML models}
\label{ssec:models}
\noindent
We first discuss how we developed the LSTM (§\ref{sssec:lstm}), then focus on the LLM (§\ref{sssec:llm}) and conclude by explaining how we integrated these models in the ATS (§\ref{sssec:combining}). Additional  technical details are provided in the Appendix~\ref{app:environment}.

\subsubsection{\textbf{Price model (LSTM)}}
\label{sssec:lstm}
The price–forecasting component is an LSTM trained on multivariate input sequences.  
For each asset $s$, the model analyses a rolling window of $W_p{=}50$ (used also in~\cite{rizvani2025ephemeral}) daily bars
(Open, High, Low, Close, Volume), $X_{t-W_p+1:t,s}$ and produces a one–day–ahead price forecast,
\begin{equation}
  \widehat{P}_{t+1,s} = f_\theta\!\big(X_{t-W_p+1:t,s}\big).
\end{equation}
From this forecast we derive a directional signal,
\begin{equation}
  \Delta_{t,s} = \frac{\widehat{P}_{t+1,s} - P_{t,s}}{P_{t,s}},
\end{equation}
which expresses how much the model expects the price of asset $s$ to rise or fall on the next day relative to today’s price.
The LSTMs are trained on observations from Jan. 2013 to Dec. 2023 (as also done in~\cite{rizvani2025ephemeral})

\subsubsection{\textbf{News model (LLM)}}
\label{sssec:llm}
The second component of the ATS is an LLM-based sentiment extractor.  
We employ FinBERT~\cite{huang2023finbert}, a transformer-based LLM pre-trained on financial corpora and fine-tuned for sentiment classification.
Before the sentiment-classification step, however, it is first needed to determine the stock $s$ which a headline refers to.\footnote{Indeed, if the ATS gets headlines from public feeds, it is possible that such an headline is unrelated to any stock in the portfolio (in which case, it must be discarded). Conversely, if the ATS uses Refinitiv, it is possible that the headline may be unrelated to the stock mentioned in the query---e.g., the headline ``Nvidia shares surge 13\%, lift market value a record \$330 billion'' appears not only when querying Refinitiv's API with ``NVDA'' but also for ``MSFT'' (because ``MSFT'' is mentioned in the text of~\cite{reuters2024nvidia}). Given that the sentiment-scoring is applied to the headline, it is fundamental to ascertain the stock mentioned in the headline (e.g., the previously mentioned headline is positive for NVDA, but the text of the corresponding news states that ``MSFT'' stocks fell by 1\%, which is a negative sentiment \textit{not captured by the headline}).} Such a task is done also by means of an LLM, for which we devise the prompt in Listing~\ref{app:prompt-map}; we manually checked \smamath{\approx}7,000 stock-name associations produced by such a prompt/LLM, and confirmed they were always correct. Nonetheless, for each headline associated with asset $s$ on day $t$, FinBERT outputs a scalar polarity score $s^{(j)}_{t,s} \in [-1,1]$ (we report the prompt in Listing~\ref{app:prompt-sent}: qualitative manual inspection on 7,000 headlines confirms that the sentiment was always predicted correctly).
We aggregated these values by computing the daily mean: $s_{t,s} = \frac{1}{N_{t,s}} \sum_{j=1}^{N_{t,s}} s^{(j)}_{t,s},$
and further stabilize the signal using a 7-day moving average: $\bar{s}_{t,s} = \frac{1}{7}\sum_{k=0}^{6} s_{t-k,s}.$
This procedure reflects industry practice where news are used to derive slow-moving sentiment indicators for trading~\cite{LSEGNewsService2025} (albeit implementation details of real-world systems are not publicly available; we will empirically validate our choices).

\subsubsection{\textbf{Integrating the models in the ATS}}
\label{sssec:combining}
We use our models (LSTM and LLM) in two ways---each representing a distinct ATS: one making its decisions only based on the output of the LSTM, and the other aggregating the output of both the LSTM and the LLM. The reason for this separation is to ensure that our ATS is implemented correctly: we are not aware of any open-source implementation of ATS that combine LSTM with LLMs. The only publicly-available ATS we are aware of is the one in~\cite{rizvani2025ephemeral}, which only uses an LSTM. So, by developing two ATS, we can ascertain if the added value of the LLM results in a more profitable ATS---which would justify its deployment in the real world.\footnote{In a sense, this experiment can be seen as an ablation study and a sanity check. If adding the LLM would yield an ATS with worse performance, then any security assessment against such an ATS would have poor validity.} We implement the two ATS as follows.
\begin{itemize}[leftmargin=*]
    \item \textit{LSTM only.} This ATS converts the output signal of the LSTM into discrete actions using a symmetric threshold $\tau$: go long if $\Delta_{t,s} > \tau$, go short if $\Delta_{t,s} < -\tau$, and otherwise hold cash. Position sizes are determined by investing a fixed capital fraction $\alpha \in (0,1]$ per active signal. The parameters $(\tau,\alpha)$ are tuned on the training set and fixed at test time. 
    
    \item \textit{LSTM+LLM.} This ATS takes the two signals (the LSTM one, and the LLM one) and fuses them into a single score:
\begin{equation}
  \Sigma_{t,s} = w_p \,\Delta_{t,s} + w_n \,\bar{s}_{t,s},
  \quad w_p,w_n \ge 0,\; w_p + w_n = 1.
\end{equation}
Such a ``hybrid'' strategy follows the same thresholding logic as the ``LSTM only'' case, but on $\Sigma_{t,s}$. Hyperparameters $(w_p, w_n, \tau, \alpha)$ are tuned on the training set and fixed at test. 
\end{itemize}
These implementations are available in our repository~\cite{ourRepo}.

\subsection{Baseline Assessment of our ATS (no-attack scenario)}
\label{ssec:baseline}
\noindent
We assess our (two) ATS in the absence of attacks. Such a preliminary assessment has a twofold goal: {\small \textit{(i)}}~ascertain that both of our ATS yield a profit, and {\small \textit{(ii)}}~ascertain that ``hybrid'' ATS yields more profits than the ``LSTM only'' ATS.

To this end, we simulate daily trading across all ten assets with portfolio-level cash management; such simulations span across 14 months (from Feb, 2024 to Apr, 2025, ensuring no overlap with the training set). We set: initial capital=\smamath{1,000,000}\$ (common~\cite{quantiacs}); transaction cost=\smamath{0.005}\$ per share~\cite{IteractiveBroker}; slippage cost=\smamath{0.02} to account for the difference between the expected and actual execution prices~\cite{lv2019empirical}. 

\begin{figure}[!t]
  \centering
  \includegraphics[width=\columnwidth]{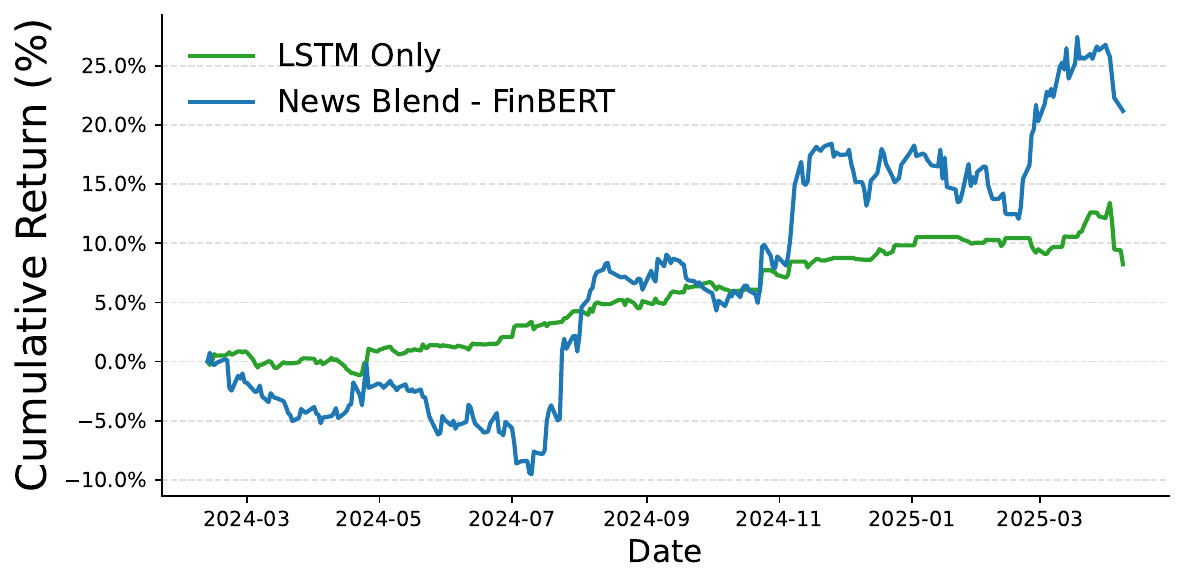} 
  \vspace{-6mm}
  \caption{\textbf{Baseline performance of the LSTM-only and LSTM+LLM ATS (no-attack case)}. The LSTM+LLM ATS is the most profitable of the two, confirming the validity of our implementation and subsequent assessment.}
  \label{fig:baseline_equity}
  \vspace{-5mm}
\end{figure}

To measure the performance of the ATS, we use the cumulative returns (CR). We report the results of the simulation in Fig.~\ref{fig:baseline_equity}. The ``LSTM only'' ATS has a stable growth of its CR, which amounts to an increase of 7.9\% at the end of the testing period. In contrast, adding the LLM leads to a much more profitable ATS, with an overall CR at the end of the testing period of 19.22\%. These results confirm that sentiment signals from financial LLMs can provide non-redundant information that enhances predictive power and trading precision under realistic constraints. Hence, we can use such an ``LSTM+LLM'' ATS for our security assessment.

%% file: sections/5-evaluation.tex
\section{Security Assessment of our Threat Model}
\label{sec:evaluation}

\noindent
We now measure the impact of our proposed manipulations. We begin by describing how we practically realised the perturbations~(§\ref{ssec:attacks}), and then present the results of the homoglyph attack (§\ref{ssec:attack1}) and of the hidden-text attack~(§\ref{ssec:attack2}),

\subsection{Attack Implementation and Evaluation}
\label{ssec:attacks}
\noindent
We first outline the common assumptions (§\ref{sssec:assumptions}), then explain our manipulation procedure (§\ref{sssec:apply}), and finally describe how we organize our security assessment (§\ref{sssec:assessment}).

\subsubsection{\textbf{Common Assumptions}}
\label{sssec:assumptions}
The attacker can manipulate the news headlines before they enter the ATS; it is implicitly assumed that such manipulations are not sanitized before reaching the ATS.
Recall that the attacker knows at least one stock $s^\star$ of the targeted ATS' portfolio. Given that our ATS has 10 stocks in its portfolio, and to simulate an attacker with limited knowledge, we assume that the attacker knows only one stock. Without loss of generality, we assume that the attacker can operate on a single day: therefore, if one (or more) news related to $s^\star$ appear on any given day, the attacker will manipulate the corresponding headline(s). Afterwards, the attacker will not attempt to apply any sort of manipulation. Hence, during the entire 14-month-long test window, the ATS will receive the manipulated headline on at most one day.

\subsubsection{\textbf{Manipulation (and proof-of-concept test)}}
\label{sssec:apply}
As a prerogative, we manipulate the headline before they reach the LLM (i.e., we apply a ``problem-space'' perturbation, albeit we do not alter real-world news!). We proceed as follows:
\begin{itemize}[leftmargin=*]
    \item For the \textit{homoglyph attack}, and for all news mentioning the stock $s^\star$ on any given day, we replace the Latin characters included in the name of stock $s^\star$ with their Cyrillic equivalent (we report in Tables~\ref{tab:mapping_char}, \ref{tab:mapping_ticker}, \ref{tab:mapping_company} the specific Latin\smamath{\Rightarrow}Cyrillic replacements). Such a process induces the LLM to fail the association. As a proof-of-concept experiment, we took nine news mentioning NVIDIA (supported by Refinitiv and included in our ATS portfolio) on August 1st, 2024 and applied our attack. The results are in Table~\ref{tab:misrouting-nvda}, showing that the LLM would normally be able to recognize eight out of nine tickers of NVDA, but after our attack it is not able to recognize a single one; the sentiment is barely affected (notice the \smamath{\Delta}). As a result, the LLM would not be able to make an informed decision on whether to do anything with NVDA stocks on that day. 

    \item For the \textit{hidden-text attack}, we follow the same logic as in the homoglyph attack (in terms of ``when to act''. However, instead of replacing characters, we add a fixed, invisible string in the HTML: {\footnotesize \texttt{<span style="display:none">losses and layoffs</span>}} Such a string should always induce a ``negative'' sentiment (meaning that if the sentiment of the headline is positive, the expectation is that the LLM will output a less-positive sentiment; and if the sentiment is already negative, it will provide an even more negative sentiment). 
    As another proof-of-concept experiment of this use case, we applied such a manipulation on the nine headlines of NVDA on August 1st, and report the output of the LLM to all such headlines in Table~\ref{tab:htmlinject-nvda}. We see that there is a change in polarity for seven out of nine news, whereas the remaining two become more negative. The daily sentiment $\bar{s}_{t,s}$ moves from positive to strongly negative, despite no visible change (from a human viewpoint). 
\end{itemize}
Aside from the above, we do not manipulate anything else. 

\subsubsection{\textbf{Evaluation procedure}}
\label{sssec:assessment}
Recall that our attacker can only apply manipulations on a single day, and that he/she only knows a single stock (out of 10) within the ATS portfolio; moreover, our test window spans across 14 months. So, for a comprehensive assessment, we evaluate \textit{all possible combinations} of these circumstances. In other words, to evaluate, e.g., the homoglyph attack and assuming that $s^\star$=NVIDIA, we simulate what happens if the attacker applies the perturbation on day-1 of the test window for news related to NVIDIA and test the corresponding effects on the ATS; then we repeat the process, but by applying the manipulation on day-2, and so on until the last day of the test window. We then repeat the process again, but by assuming the stock known by the attacker is a different one (e.g., $s^\star$=GOOGL). We continue until we exhaust all the possibilities, which are given by: 10 (stocks) * 420 (days) *2 (attacks). Hence, overall, we test \smamath{\approx}8k perturbations. For each of these, we measure the CR at the end of the test window and compare it with the baseline CR, and we also log additional details. The entire evaluation procedure can be formally expressed as follows. Let $\mathcal{C}=\{(t,s):$ be at least one headline for $s$ on day $t\}$. For each $(t^\star,s^\star)\in\mathcal{C}$ we: {\small \textit{(i)}}~run a clean backtest; {\small \textit{(ii)}}~apply the manipulation \emph{only} to $\mathcal{H}_{t^\star,s^\star}$ and rerun under identical market data, costs, and execution rules; {\small \textit{(iii)}}~record model effects (routing failures for homoglyphs, $\Delta s_{t^\star,s^\star}$ for hidden text) and system effects (action flip $\mathbb{I}\{\tilde{a}_{t^\star,s^\star}\neq a_{t^\star,s^\star}\}$ and portfolio delta $\Delta\mathrm{CR}_{t^\star,s^\star}=\!\tilde{\mathrm{CR}}-\mathrm{CR}$ (where $\tilde{\mathrm{CR}}$ denotes the cumulative returns of the ATS under attack). 
We then aggregate across $\mathcal{C}$ to report flip rates, and the distributions of $\Delta s_{t,s}$ and $\Delta\mathrm{CR}_{t,s}$, with stratification by ticker and news volume.

\begin{table}[!htbp]
\centering
\caption{Homoglyph misrouting (FinBERT, NVIDIA, 1\textsuperscript{st} Aug 2024).}
\label{tab:misrouting-nvda}
\vspace{-2mm}
\scriptsize
\setlength{\tabcolsep}{3pt}
\begin{tabularx}{\columnwidth}{@{}r l l *{3}{>{\centering\arraybackslash}X} c@{}}
\toprule
\textbf{\#} & \textbf{Clean map} & \textbf{Attack map} & \textbf{Clean Sent.} & \textbf{Attack Sent.} & $\boldsymbol{\Delta}$ & \textbf{Flip} \\
\midrule
1 & NVDA   & Unrecognized &  0.713 &  0.787 & +0.075 & F \\
2 & NVDA   & Unrecognized &  0.854 &  0.851 & -0.003 & F \\
3 & Unrec. & Unrecognized &  0.908 &  0.908 &  0.000 & F \\
4 & NVDA   & Unrecognized &  0.035 &  0.029 & -0.006 & F \\
5 & NVDA   & Unrecognized &  0.832 &  0.808 & -0.024 & F \\
6 & NVDA   & Unrecognized &  0.832 &  0.808 & -0.024 & F \\
7 & NVDA   & Unrecognized & -0.290 & -0.210 & +0.080 & F \\
8 & NVDA   & Unrecognized & -0.290 & -0.068 & +0.223 & F \\
9 & NVDA   & Unrecognized &  0.901 &  0.918 & +0.017 & F \\
\bottomrule
\end{tabularx}
\end{table}

\begin{table}[!htbp]
\centering
\scriptsize
\caption{Hidden-text manipulation (FinBERT, NVIDIA, 1\textsuperscript{st} Aug 2024): sentiment per headline.}
\label{tab:htmlinject-nvda}
\vspace{-2mm}
\setlength{\tabcolsep}{3pt}
\begin{tabularx}{\columnwidth}{@{}r *{3}{>{\centering\arraybackslash}X} c@{}}
\toprule
\textbf{Headline} & \textbf{Clean} & \textbf{Attack} & $\boldsymbol{\Delta}$ & \textbf{Flip} \\
\midrule
1 &  0.901 & -0.916 & -1.817 & T \\
2 &  0.832 & -0.939 & -1.771 & T \\
3 &  0.832 & -0.939 & -1.771 & T \\
4 &  0.854 & -0.883 & -1.737 & T \\
5 &  0.713 & -0.936 & -1.649 & T \\
6 &  0.908 & -0.558 & -1.466 & T \\
7 &  0.035 & -0.945 & -0.980 & T \\
8 & -0.290 & -0.961 & -0.670 & F \\
9 & -0.290 & -0.961 & -0.670 & F \\
\bottomrule
\end{tabularx}
\end{table}

\subsection{Assessment of the Homoglyph Attack}
\label{ssec:attack1}
\noindent
We present the results of the homoglyph attack on the system-wide ATS at three different granularity levels: ``one day, one stock'', ``all days, one stock'', and ``all days, all stocks''.

\begin{figure}[!t]
  \centering
  \includegraphics[width=\columnwidth]{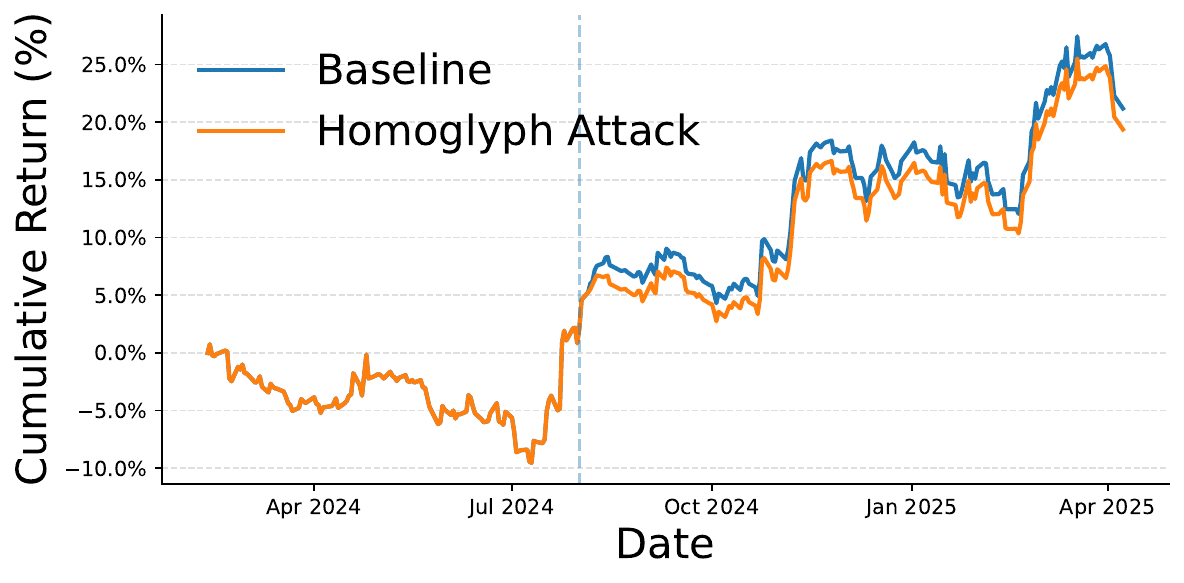}
  \vspace{-7mm}
  \caption{CR of the ATS under clean vs.\ homoglyph (1\textsuperscript{st} Aug 2024, NVIDIA).}
  \label{fig:misrouting-cr}
  \vspace{-5mm}
\end{figure}

\textbf{One day [August 1st, 2024], one stock [NVIDIA].} We report in Fig.~\ref{fig:misrouting-cr} the impact on the cumulative returns of the ATS caused by the attack we discussed in §\ref{sssec:apply} (i.e., on August 1st, 2024, affecting nine headlines of NVIDIA). We can see a slight decrease of the CR. Importantly, however, the ATS \textit{still generates a profit}. This is crucial: if the ATS stopped being profitable, its owners would stop using the ATS. What makes such an attack subtle is precisely this aspect: the owners would not notice that they are being attacked, because there is no perceivable indicator... and yet, their ATS is yielding less money due to an incorrect decision made as a consequence of the manipulation that occurred on August  1st, 2024, which led to a cascading effect and a drastic reallocation of resources, affecting all future trading decisions of the ATS.

\textbf{All days, one stock [NVIDIA].} The previous results assumed the attack took place on a single day. However, the attacker has over 400 days in which they could theoretically apply the manipulation on NVIDIA-related news. Here, we assess what happens across this entire test window: the goal is showing, on average, how much the ATS would lose if the attacker randomly chose one day to apply their manipulations on NVIDIA-related headlines. From our results, we obtain: $\mathrm{CR}=19.22$ and $avg(\tilde{{\mathrm{CR}}})\approx 17.5\%$.
In other words, a single-day homoglyph overlay on NVDA reduces end-of-window performance by about $1.7$ percentage points on a \$1M book under identical execution, which is equivalent to a net loss of 17,200\$ (given by $\Delta\$ \;=\; I_0 \cdot \Delta\mathrm{CR}_{t^\star,s^\star}$, where $I_0=1M\$$).

\textbf{All days, all stocks.} We can use the same accounting to aggregate the results across all days and for all stocks of the portfolio. We found that, against the specific LLM (i.e., FinBERT), 5,957 out of 6,012 headlines (i.e., 99.1\%) misled the stock-name association. The impact of such misroutings on the overarching ATS is reported in Table~\ref{tab:system_level_summary} (showing the impact of the homoglyph attack against the entire ATS). The results are fascinating: the worst possible scenario for the ATS would be if the attacker applied the manipulation on headlines of TSLA on February 28th, 2024, which would lead to a drop of 17.7 percentage points w.r.t. the baseline CR. Intriguingly, however, the attacker can also ``lose'': if the manipulation is applied on February 29th, 2024 on NVIDIA, the ATS would yield a higher CR w.r.t. the baseline. Regardless, on average, and by aggregating all possible combinations, the homoglyph manipulation induce a drop to the CR of 3.67\%. Notably, the ATS always made a different decision as a result of this attack. 

\begin{table}[!htbp]
\centering
\caption{System-level impact of Unicode homoglyph substitution (FinBERT). Each trial attacks exactly one stock–day; others remain clean. $\Delta$CR in percentage points relative to paired clean run.}
\label{tab:system_level_summary}
\vspace{-2mm}
\scriptsize
\setlength{\tabcolsep}{3pt}
\begin{tabularx}{\columnwidth}{@{}l >{\centering\arraybackslash}X@{}}
\toprule
\textbf{Metric} & \textbf{Value} \\
\midrule
Days with decision change [\%]   & 100.0 \\
Mean $\Delta$CR [pp]           & -3.67 \\
Mean $|\Delta$CR| [pp]         &  4.46 \\
Worst case (TSLA, 28 Feb 2024) & -17.70 \\
Best case (NVDA, 29 Feb 2024)  &  +4.58 \\
\bottomrule
\end{tabularx}
\end{table}

\begin{cooltextbox}
\textbf{\textsc{Takeaway.}} Homoglyph edits leave headlines visually unchanged but break stock–name association. About \text{99\%} of edited headlines fail to map to the correct ticker, thinning the day’s sentiment. In paired backtests, \textit{all} attacked days produce at least one action flip. Portfolio impact averages about \text{-3.7 pp}, with occasional double-digit losses, under identical prices, costs, and execution.
\end{cooltextbox}

\subsection{Assessment of the Hidden-text Attack}
\label{ssec:attack2}
\noindent
This section has the same structure as the previous one (§\ref{ssec:attack1}), so we simply report the results and analyse them.

\begin{figure}[!htbp]
\vspace{-2mm}
  \centering
  \includegraphics[width=\columnwidth]{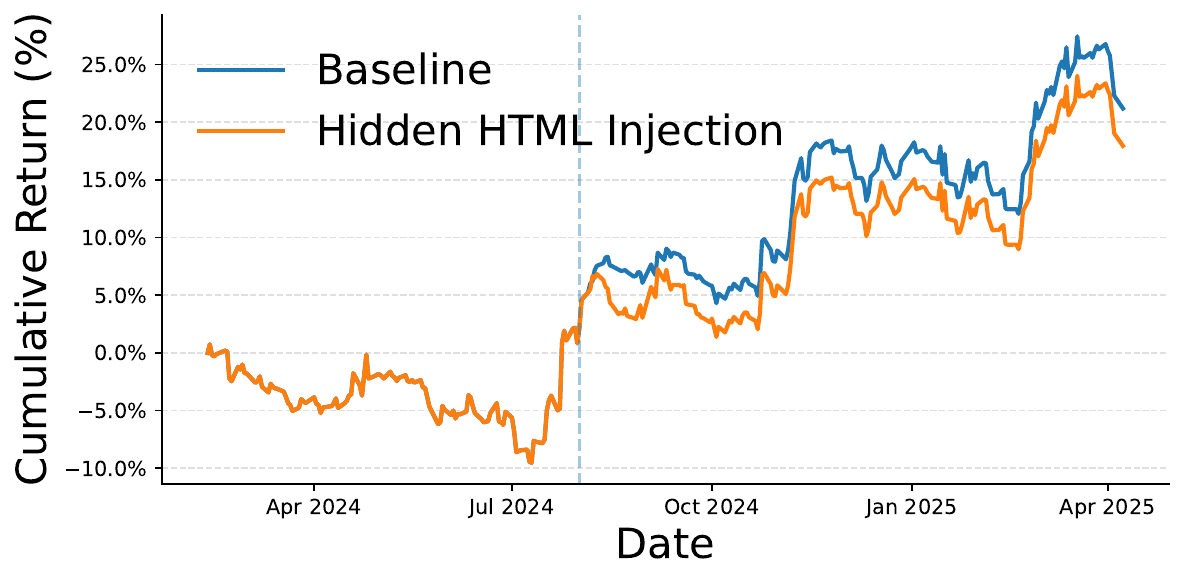}
  \vspace{-7mm}
  \caption{CR of the ATS under clean vs.\ hidden-text (1\textsuperscript{st} Aug 2024, NVIDIA).}
  \label{fig:htmlusecase}
  \vspace{-2mm}
\end{figure}

\textbf{One day [August 1st, 2024], one stock [NVIDIA].} The impact on the CR of the hidden-text attack is shown in Fig.~\ref{fig:htmlusecase}. This case also leads to a lower CR of the ATS w.r.t. the baseline, which is even more prominent than that induced by the homoglyph attack (which affected exactly the same news, and exactly the same day). This is because the ATS made a substantially wrong decision on this day due to the (faked) highly-negative sentiment perceived by the manipulated headlines---which no human noticed.

\textbf{All days, one stock [NVIDIA].} Accounting for the entire testing window, the attack leads to $avg(\tilde{{\mathrm{CR}}})\approx 16.0\%$. Thus, on average, a single-day hidden HTML overlay on NVDA decreases end-of-window performance by about $3.2$ percentage points, corresponding to an approximate \$32.2k shortfall on a \$1M initial capital investment.

\textbf{All days, all stocks.} We can derive the aggregated effects of the hidden-text attacks. Specifically:
\begin{itemize}[leftmargin=*,noitemsep,topsep=0pt]
  \item Model level: 65.6\% of headlines flip sentiment polarity.
  \item System level: 77.3\% of trials reduce ATS CR; mean effect $-3.18$\,pp $\Delta$CR; worst case $-17.7$\,pp (TSLA, 28 Feb 2024).
\end{itemize}
Interestingly, the ``worst-case'' falls on exactly the same day (February 28th, 2024), and affects exactly the same stock (TSLA), as in the homoglyph attacks (causing also the same drop in the CR, i.e., 17.7\%).

\begin{cooltextbox}
\textbf{\textsc{Takeaway.}} Hidden-text affects sentiment without altering what humans see: headlines are visually unchanged, yet the LLM parses the hidden text. In paired backtests this induces action flips and, on average, a multi--percentage-point drop in portfolio CR, with occasional double-digit losses (max: -17.7\%), under identical prices, costs, and execution.
\end{cooltextbox}

%% file: sections/6-cross-model.tex
\section{Cross-model Evaluation (Transferability)}
\label{sec:cross}
\noindent
To assess the transferability of our attack against ATS powered by LLMs different from FinBERT, we expand our assessment by considering 9 different LLMs. 
We describe our setup where we assess the baseline performance of each different ATS variant (§\ref{ssec:setup}), then we evaluate both attacks at the model level (\ref{ssec:model}) and at the system level~(§\ref{ssec:system}).

\subsection{Baseline Performance of the new ATSs}
\label{ssec:setup}
\noindent
To ensure broad coverage, we consider different families of LLMs. Specifically: two additional finance-oriented models (FinGPT, FinLLaMA) and six general-purpose models (O3, O3 Pro, 4o, 4o-mini-high, 4o-mini, GPT-5, Gemini Pro 1.5). 

\begin{table}[!b]
\centering
\vspace{-4mm}
\caption{Baseline portfolio performance (clean runs; no attack).}
\label{tab:baseline-llm}
\vspace{-2mm}
\renewcommand{\arraystretch}{1.2}
\resizebox{\columnwidth}{!}{
\begin{tabular}{c|*{10}{c}}
\toprule
 & \rotatebox{40}{FinLLama} 
 & \rotatebox{40}{FinBERT} 
 & \rotatebox{40}{FinGPT} 
 & \rotatebox{40}{O3} 
 & \rotatebox{40}{O3 Pro} 
 & \rotatebox{40}{4o} 
 & \rotatebox{40}{4o-mini-high} 
 & \rotatebox{40}{4o-mini} 
 & \rotatebox{40}{GPT-5} 
 & \rotatebox{40}{Gmn.Pro 1.5} \\
\midrule
CR [\%] & 
19.5 & 19.2 & 17.0 & 16.8 & 16.5 & 16.3 & 16.1 & 15.9 & 15.8 & 15.2 \\
\bottomrule
\end{tabular}
}
\vspace{-2mm}
\end{table}

Hence, we take our original ATS (discussed in §\ref{sec:ats}) and craft a different variant by replacing FinBERT with any of the aforementioned models. We then assess the baseline performance of each new ATS by measuring the corresponding CR over the same time period. The results are shown in Table~\ref{tab:baseline-llm} (we also report the CR for FinBERT). 

We see that FinLLaMa is the model that yields the highest profit (the CR increases by 19.5\%), whereas Gemini Pro 1.5 yields the lowest profits (the CR increases by 15.2\%). These results are expected: LLMs focused for the financial domain provide higher revenues than general-purpose models.

\subsection{Attacking multiple ATS (Model-only Evaluation)}
\label{ssec:model}

\noindent
We now launch our attacks against the new ATSs. We begin by assessing the impact of our attacks against the specific LLMs.

We report the results of the Homoglyph attack in Table~\ref{tab:llm-misrouting}, showing the mapping accuracy before/after homoglyph spoofing. The baseline accuracy of all models degrades substantially for most models. Particularly, the finance-specific LLMs are highly affected (almost 80\% drop). Perhaps surprisingly, O3 appears quite robust (only 8.32\% drop), even more than O3 Pro (which suffered a 5x larger drop).

Put simply, aside from perhaps O3, all these models cannot recognize the stock in the presence of an homoglyph attack.

\begin{table}[!t]
\centering
\caption{Model-level robustness to homoglyph misrouting (mapping acc).}
\label{tab:llm-misrouting}
\vspace{-0mm}
\scriptsize
\setlength{\tabcolsep}{3pt}
\begin{tabularx}{0.7\columnwidth}{@{}l
  S[table-format=2.2]
  S[table-format=2.2]
  S[table-format=2.2]@{}}
\toprule
\textbf{Model} & {\textbf{Clean [\%]}} & {\textbf{Attack [\%]}} & {\textbf{Impact [pp]}} \\
\midrule
O3              & 96.72 & 88.40 &  8.32 \\
GPT-5           & 94.50 & 50.00 & 44.50 \\
O3 Pro          & 95.02 & 46.43 & 48.59 \\
4o              & 94.91 & 35.98 & 58.93 \\
4o-mini-high    & 94.38 & 35.65 & 58.73 \\
4o-mini         & 89.38 & 14.10 & 75.28 \\
Gemini Pro 1.5  & 91.27 & 12.05 & 79.22 \\
FinGPT          & 93.21 & 12.80 & 80.41 \\
FinLLaMA        & 90.87 & 11.17 & 79.70 \\
FinBERT         & 90.04 & 0.90 & 89.14 \\
\bottomrule
\end{tabularx}
\vspace{-1mm}
\end{table}

Table~\ref{tab:llm-html} reports sentiment flips for hidden-text edits; most models parse the invisible clause and shift sentiment, but the effectiveness varies across models. O3 shows the lowest flip rate (40\%), followed by GPT-5 (43\%); most others exceed 80\%. Thus, in our setting O3 is the most resilient to invisible-clause attacks by frequency. However, when O3 does get flipped, the average shift is large ($\Delta$sentiment $=-0.90$), indicating rare-but-strong effects. Several models, including O3, also exhibit higher confidence as a result of the attack (e.g., O3: $+0.31$), so confidence-based monitoring alone may miss failures. Overall, hidden text remains a reliable adversarial tactic, especially against the finance-tuned LLMs. 

\begin{table}[!htbp]
\centering
\caption{Model-level robustness to hidden-text manipulation (flip rate and deltas). \emph{Flip rate} counts only polarity reversals ($+1\leftrightarrow-1$)}
\label{tab:llm-html}
\vspace{-2mm}
\small
\setlength{\tabcolsep}{3pt}
\begin{tabularx}{\columnwidth}{@{}l *{3}{>{\centering\arraybackslash}X}@{}}
\toprule
\textbf{Model} & \textbf{Flip rate [\%]} & $\Delta$\textbf{Sentiment} & $\Delta$\textbf{Confidence} \\
\midrule
O3              & 40.0 & -0.90 & +0.31 \\
GPT-5           & 43.0 & -0.10 & +0.12 \\
4o              & 45.0 & -0.04 & +0.18 \\
4o-mini         & 46.0 & +0.02 & -0.04 \\
4o-mini-high    & 80.7 & -0.99 & +0.45 \\
O3 Pro          & 86.4 & -1.01 & +0.26 \\
FinBERT         & 65.6 & -1.03 &  0.00 \\
FinGPT          & 84.0 & -0.31 &  0.00 \\
FinLLaMA        & 85.0 & -0.45 & -0.05 \\
Gemini Pro 1.5  & 86.0 & -0.52 & -0.08 \\
\bottomrule
\end{tabularx}
\end{table}

\subsection{System-wide Evaluation}
\label{ssec:system}
\noindent
We propagate manipulated headlines through the ATS, one stock–day at a time, and measure the impact of the attack.

Table~\ref{tab:system-cross-homoglyph} focuses on the homoglyph attack against the system-wide ATS. As expected, O3 is the most robust model (verified with a t-test, \smamath{p<.05}: the overall drop in CR is of only 0.5\%; across the 14-months test window, the attacks launched in 91\% of the days had no effect on O3). This is because O3 is a powerful reasoning model which was able to recognize, and automatically sanitize, the presence of cyrillic characters. However, the other LLMs suffered substantial impact. GPT-5 had an average drop of 2\% and 55\% of the manipulations affected it. The most vulnerable models are 4o-mini, Gemini Pro 1.5, and the finance-specific LLMs: an attacker would be able to reliably induce economic losses against these, given that the attack was successful for over 75\% of the days within the test window (leading to sensible drops in the CR).

\begin{table}[!htbp]
\centering
\caption{System-level impact of homoglyph misrouting across models.}
\label{tab:system-cross-homoglyph}
\vspace{-2mm}
\small
\setlength{\tabcolsep}{3pt}
\begin{tabularx}{\columnwidth}{@{}l *{3}{>{\centering\arraybackslash}X}@{}}
\toprule
\textbf{Model} & \textbf{Days impacted [\%]} & \textbf{Mean $|\Delta$CR| [pp]} & \textbf{Worst case [pp]} \\
\midrule
O3              & 8.3  & 0.5 & -2.0  \\
GPT-5           & 44.5 & 2.0 & -7.0  \\
O3 Pro          & 48.6 & 2.2 & -8.0  \\
4o              & 58.9 & 2.6 & -10.0 \\
4o-mini-high    & 58.7 & 2.6 & -10.0 \\
4o-mini         & 75.3 & 3.5 & -13.5 \\
Gemini Pro 1.5  & 79.2 & 3.8 & -15.0 \\
FinGPT          & 80.4 & 4.0 & -16.0 \\
FinLLaMA        & 79.7 & 3.9 & -16.0 \\
FinBERT         & 100.0 & 4.46 & -17.70 \\
\bottomrule
\end{tabularx}
\vspace{-2mm}
\end{table}

The effects of the hidden-text manipulation are reported in Table~\ref{tab:system-cross-html}. The trend is similar to that shown in Table~\ref{tab:system-cross-homoglyph}, with O3 being significantly more robust (verified with a t-test, \smamath{p<.05}); whereas the three finance-specific LLMs, alongside Gemini Pro 1.5 and 4o-mini are the most vulnerable.

\begin{table}[!htbp]
\centering
\caption{System-level impact of hidden-text manipulation across models.}
\label{tab:system-cross-html}
\vspace{-2mm}
\small
\setlength{\tabcolsep}{3pt}
\begin{tabularx}{\columnwidth}{@{}l *{3}{>{\centering\arraybackslash}X}@{}}
\toprule
\textbf{Model} & \textbf{Days Impacted [\%]} & \textbf{Mean $|\Delta$CR| [pp]} & \textbf{Worst case [pp]} \\
\midrule
O3              & 9.1  & 0.7 & -2.4  \\
GPT-5           & 46.8 & 2.1 & -7.6  \\
O3 Pro          & 51.2 & 2.4 & -8.5  \\
4o              & 60.4 & 2.7 & -10.8 \\
4o-mini-high    & 61.0 & 2.8 & -11.2 \\
4o-mini         & 76.1 & 3.6 & -14.0 \\
FinBERT         & 77.3 & 3.2 & -17.7\\
Gemini Pro 1.5  & 80.3 & 3.9 & -15.4 \\
FinGPT          & 81.2 & 4.1 & -16.3 \\
FinLLaMA        & 80.6 & 4.0 & -16.1 \\
\bottomrule
\end{tabularx}
\vspace{-4mm}
\end{table}

\begin{cooltextbox}
\textbf{\textsc{Takeaway.}}
These vulnerabilities are not unique to FinBERT. Finance-specific and general-purpose LLMs alike {\small \textit{(i)}}~mis-handle mixed-script stock names and {\small \textit{(ii)}}~parse hidden HTML clauses; even a single-day manipulation can shift portfolio outcomes. However, O3 appears to be robust.
\end{cooltextbox}

%% file: sections/7-survey.tex
\section{Survey with FinTech Professionals}
\label{sec:survey}
\noindent
To validate our research design, we carried out an original user study with practitioners in the FinTech sector. We first outline our methodology (§\ref{ssec:methodology}), then describe our sample (§\ref{ssec:sample}) and finally present our findings (§\ref{ssec:survey_findings}).

\subsection{Methodology}
\label{ssec:methodology}
\noindent
Despite abundant research suggesting that integrating LLMs in ATS is sensible~\cite{li2025enhancing}, real-world evidence that this is indeed happening is scarce. In other words: do FinTech practitioners truly use LLMs to make (automated) trading decisions? To address this question, which serves to scrutinize whether our threat model (and, hence, experimental evaluation) depicts a realistic scenario, we devised an online questionnaire.

\textbf{Questionnaire} Our questionnaire was created via Google Forms, and has a total of 13 questions. After introducing the participant to the purpose of our survey and asking (Q1) for their permission to use their data for research, we ask two job-related questions (Q2: ``which industry describes your organization?'' and Q3: ``what is your role?''). Then, we ask eight closed questions. Among these, the four most relevant ones are:
(Q4) ``In your industry, how common is the use of AI or LLMs for sentiment analysis or news-driven trading decisions?''; 
(Q6) ``What do you use  AI/LLMs for?''
(Q7) ``Who prepares news data for use in analytics or trading models?''
(Q9) ``How realistic do you consider the following scenario: `Manipulated or malicious financial news mislead an AI/LLM system and results in an incorrect trading decision.'?''
We then ask two open-text questions, inquiring if there is anything they want to add, and to provide an email address for data deletion and/or receive further insights about our research. We report the questionnaire (verbatim), alongside the possible answers to each question, in our repository~\cite{ourRepo}.

\textbf{Dissemination.} We recruited our participants via convenience sampling~\cite{emerson2015convenience,antoun2016comparisons} (as also done, e.g., in~\cite{schroer2025sok}). Since we were interested in FinTech professionals, we privately reached out to practitioners in the field that we found via OSINT~\cite{glassman2012intelligence} and also on social networks (e.g., LinkedIn). Such an approach ensured that only qualified individuals would participate in our survey; we did not know what these individuals would have answered to our questions beforehand. Overall, we sent 83 invitations within Sept. 10--23, 2025 (i.e., \textit{after} our evaluation, making our survey a fair way to validate our choices).

\textbf{Ethics.} Our institutions do not mandate a formal IRB approval to carry out such a user study. Yet, we followed established ethical guidelines~\cite{bailey2012menlo}. Participants were informed of the nature of our study and willingly gave their informed consent. We do not inquire for any sensitive or personally-identifiable information~\cite{sensitiveEU,sensitiveUSA}. The questionnaire is anonymous (even if we privately contacted potential participants, by default we do not know who filled the questionnaire). We did not use deception and offered no compensation to participate in the survey (which helps avoiding bias for fast responses). Participants could withdraw at any point in time, and we offered the possibility to delete the data we collected (participants know our identities). We measured the time to fill the questionnaire in 10 minutes, so participants could not suffer any sort of harm by participating in our user study.

\subsection{Sample Description}
\label{ssec:sample}
\noindent
We received 27 valid responses (response rate=32\%). We checked them and found no reason to believe that the questions were answered in a dishonest way. Hence, to the best of our knowledge, our survey is among the ``largest'' in related research (e.g., the study in~\cite{rizvani2025ephemeral} only has 7 participants), 

Our 27 participants pertain to organizations spanning a variety of industries. The three most popular ones are: ``Asset management'' (12, 44\%), ``Hedge fund'' (8, 30\%); ``Market/Platform Provider'' (3, 11\%). As for the role, the most popular is ``Quant Researchers'', selected by 14 (51\%), followed by ``portfolio manager'' and ``risk manager'' (both with 2 votes). All other options (e.g., ``investment specialist'') were marked once. Complete results are in our repository~\cite{ourRepo}. 

\subsection{Findings}
\label{ssec:survey_findings}
\noindent
Let us focus on the questions most relevant for our research (the complete results are provided in our repository~\cite{ourRepo}).

First, for Q4, two (7\%) stated that AI/LLMs are ``never used'' for sentiment analysis or news-driven trading decisions; two (7\%) were ``not sure,'' and four (15\%) stated it is ``rare''. In contrast, 71\% say it is ``somewhat common'' or ``common'' (both 22\%), and most (26\%) say it is ``very common''. 

Second, for Q6, 9 (33\%) stated that AI/LLMs are (or plan to be) used for ``News Sentiment Scoring''; another 9 (33\%) chose ``headline classification'', and another 9 (33\%) marked ``signal generation''. Many also selected purposes that have little to do with finance (such as ``research summarization'' and ``internal tooling/chat'', marked by 13 and 12 participants).

Third, for Q7, the majority (16, 59\%) stated that ``Data providers (e.g., Bloomberg, Refinitiv)'' are those who prepare news data for use in analytics or trading; 10 (37\%) answered with ``in-house team'', 8 (30\%) with ``shared responsibility'', 3 (11\%) with ``trading platform/broker'' and two were not sure.

Fourth, for Q9, most (55.6\%) believe that it is ``somewhat realistic'' that manipulated/malicious financial news may mislead an AI/LLM system and result in an incorrect trading decision; only 2 (7\%) believe it is ``not realistic'', whereas one is not sure, and 33\% believe it is ``very realistic''.

\vspace{-2mm}
\begin{cooltextbox}
    \textbf{\textsc{Takeaway.}} Based on our findings, we can hence argue that:\footnote{Note: we refrain from making generalizable claims from our user study, given the limited sample size. Yet, it is factual that the viewpoint of \textit{some} practitioners supports the scenario evaluated in our work.} 
    {\small \textit{(i)}}~our choice of ATS, which combines well-known LSTM for price forecasting with LLMs for news sentiment analysis, is realistically sensible;
    {\small \textit{(ii)}}~our decision to fetch news from Refinitiv is also justified;
    {\small \textit{(iii)}}~our threat model is perceived, by our practitioners, as a potential risk.\vspace{-2mm}
\end{cooltextbox}

%% file: sections/8-discussion.tex
\section{Discussion and Real-world Impact}
\label{sec:discussion}
\noindent
We distill lessons learned (§\ref{ssec:lessons}), limitations (§\ref{ssec:limitations}), and examine our threat model under a real-world lens (§\ref{ssec:applicability}).

\subsection{Lessons Learned}
\label{ssec:lessons}
\noindent
We derive three major lessons learned from our research.

First, we found that the introduction of imperceptible changes in the headlines of financial news induces LLM-driven ATS to yield a reduced profit to its owners. To make things worse, similar attacks do not require sophisticated knowledge/capabilities on the targeted ATS: even by making a change on \textit{just one headline on a randomly chosen day}, an attacker can cause a substantial loss to any entity that feeds its LLM-driven ATS with such ``adversarial news.''

Second, such a vulnerability affects a broad range of LLMs---albeit some are more robust than others. However, and crucially, LLMs that are more robust against our attack (i.e., GPT O3) are not those that yield the best cumulative returns in the absence of attacks; conversely, LLMs that yield the best profits on ``clean'' news (i.e., FinLLaMa and FinGPT) are those that are affected the most by our attack.

Third, the aforementioned findings were \textit{only possible by adopting a system-wide view}. Prior work only assessed whether LLMs could be affected by ``adversarial perturbations'' (including homoglyph attacks~\cite{cooper2025lies}). However, such an approach cannot be used to quantify the impact that such attacks have on full-fledged systems. Some models can be very robust (e.g., GPT-O3 still maintains an accuracy of 88.4\% under attack) but they still lead to financial losses if attacked. 

\subsection{Scope, Limitations, and Threat to Validity}
\label{ssec:limitations}
\noindent
First, our goal was scrutinizing the impact of news headlines manipulation against LLM-driven ATS. To our knowledge, such a threat model had never been explored in the financial context. However, we acknowledge that similar approaches (e.g., homoglyph attacks~\cite{cooper2025lies}) are known to affect LLMs.

That said, there are many ways to develop an LLM-driven ATS that uses headlines for trading decisions. For instance, one can change the frequency of the trades, or use different historical windows, or different brokers, or models trained over different datasets. \textbf{We therefore do not claim generality of our results.} By publicly sharing our resources, future work can replicate our attacks in different setups and gauge the extent to which similar attacks can affect different ATS.

To develop our ATS, we relied on the open-source framework in~\cite{rizvani2025ephemeral} which we enhanced by integrating additional modules based on prior work~\cite{akhila2025hybrid,bhat2024stock}. Our user study validated our design choices (§\ref{ssec:survey_findings}). However, real-world ATS may behave differently. Importantly: while our simulations showed that our ATS would yield a profit to its owners, we do not recommend readers to use our ATS to make actual trades!

Finally, there are also infinite ways to conceive our proposed attack. For instance, attackers can use different homoglyphs, or target different headlines. However, our evaluation is by no means small-scale (i.e., our cross-model assessment encompassed nine different, and popular, LLMs---including the very recent GPT-5) and it is factual that our attacks do cause our considered ATS to yield a lower profit. Hence, we do not see any threat to the validity of our conclusions.

\subsection{Real-world Applicability of our Threat Model}
\label{ssec:applicability}
\noindent
Is our threat model realistic? Let us discuss this question by examining the assumptions of our threat model.

First, for our attack, we assume that our manipulations reach the LLM. For instance, potential ``homoglyphs'' are not sanitized, and neither are occurrences of ``invisible text''. We argue that such an assumption is realistic, because achieving a ``perfect'' sanitization is practically hard. Let us explain.
\begin{itemize}[leftmargin=*]
    \item There are many ways to realize homoglyphs beyond replacing Latin with Cyrillic characters. As an example, an attacker can replace a lowercase L (``l'') with a pipe symbol (``|''), thereby turning ``Google'' into ``Goog|e'': sanitization attempts that indiscriminately replace all ``|'' with ``l'' may \textit{also affect benign headlines} (e.g., we found one such headline in Refinitiv's data: ``Newscasts - J.P. Morgan | Software: This Week in Earnings: PRGS, BRZE and CXM'').
    \item Hidden text attacks can also be implemented in various ways. Certain HTML methods, such as \textit{innerText}, may be able to clean cases of {\scriptsize \texttt{style="display:None"}}; yet, such methods do not work against {\scriptsize \texttt{style="font-size:0pt"}}. An attacker may even set the text color to the background color, potentially with minimal variations (e.g., for white, setting \#FFFFE instead of \#FFFFF) which would still be (nearly) illegible by humans while being processed by machines. 
\end{itemize}
We are not aware of whether news vendors (such as Refinitiv) or owners of ATS for news ingestion do sanitize the inputs. However, it is factual that input-sanitization mechanisms present tradeoffs, and ``creative'' attackers may still circumvent them, explaining why our assumption is sensible.  

Second, in our threat model, we assume that the attacker can manipulate the headline. We hypothesized that the manipulation can take place in various steps outside the ATS (see §\ref{ssec:attacker}). We envisage that the likelihood that the manipulation occurs ``inside'' a news vendor (such as Refinitiv) to be unlikely (while not strictly impossible). However, the other cases are more likely (e.g., even~\cite{boucher2022bad} hypothesized a similar scenario). We believe that, given the potential damage that such manipulation can cause (as shown by our experiments), malicious actors may resort to such tactics. Especially because, even if someone claims that, e.g., a certain news provider released an ``adversarial news'', the accused party can claim an honest mistake. Regardless, these entities are typically protected by explicit terms. For instance, in the case of Refinitiv, its Terms of Service~\cite{refinitivtos} state {\small ``WE DO NOT WARRANT OR REPRESENT THAT THE PRODUCTS OR SERVICES WILL BE DELIVERED FREE OF ANY INACCURACIES, INTERRUPTIONS, DELAYS, OMISSIONS OR ERRORS, OR THAT ANY OF THESE WILL BE CORRECTED.''}. Note: we are not claiming that Refinitiv, or any news source, may launch attacks such as the one discussed in this work. We are merely pointing out that malicious actors can introduce their ``perturbations'' in various steps of the ``news stream'' before they reach the ATS.

%% file: sections/9-mitigations.tex
\section{Countermeasures and Mitigations}
\label{sec:mitigations}

\noindent
To mitigate the impact of similar attacks, we discuss defenses (§\ref{ssec:defense}), identify (and warn) vulnerable real-world platforms affected by our considered vulnerability~(§\ref{ssec:disclosure}), and provide recommendations for practitioners (§\ref{ssec:recommendations}).

\subsection{Defenses}
\label{ssec:defense}
\noindent
We elaborate on possible defensive mechanisms against attacks stemming from our threat model.

\textbf{Post-hoc Detection.} In our experiments, we assumed the attack occurs on at most one day. However, in practice, a given ATS can be targeted by adversarial news multiple times. Ideally, one way to counter repeated occurrences is via detection and reaction mechanisms. However, there is an issue: \textit{how can one detect what cannot be perceived?} Indeed, our manipulations do not disrupt the ``availability'' of the ATS: the ATS still trades and can still close the day with a profit. What is lost is the \emph{counterfactual} margin---the extra return that would have been realized under clean inputs. Even if operators notice that the ATS is less profitable, determining that the culprit is an adversarial news (and not just a byproduct of the chaotic and unpredictable stock market) is challenging.

\textbf{Robust LLMs.} Damage mitigation can be achieved via LLMs that are intrinsically more robust to adversarial news; potentially, this can be implemented by devising specific prompts that induce the LLM to critically examine the input. For instance, in our experiments, the ``reasoning'' O3 was more robust than other LLMs. Yet, O3 was not the best LLM in the no-attack case. Therefore, integration of robust LLMs should account for the tradeoff between adversarial risk and normal revenue. Measuring the counterfactual loss of a potential attack can be used to better guide such operational decisions.

\textbf{Prevention via input sanitization.} While perfect sanitization may be unfeasible, it is still possible to counter the specific adversarial news considered in our evaluation. As a proof-of-concept, we have devised an input sanitization module that, after acquiring a headline, it checks for the presence of cyrillic-latin homoglyphs and cleans them (i.e., a reverse mapping of Table~\ref{tab:mapping_char}). Such a plugin can be deployed right before the LLM for stock-name association processes the input. We tested our module: it always defuses our homoglyph attack, and it also has a negligible overhead (processing a single headline takes \smamath{<}0.1s on a COTS system). However, as we argued in §\ref{ssec:applicability}, attackers can use other homoglyphs so our plugin cannot cover the entire attack surface. Our plugin (and corresponding evaluation) is provided in our repository~\cite{ourRepo}.

\subsection{Vulnerable Platforms (and Responsible Disclosure)}
\label{ssec:disclosure}
\noindent
We have reason to believe that our envisioned threat model can impact also operational platforms used for ATS-related purposes. Recall that our considered attack is rooted on the fact that LLMs receive, as input, data (i.e., the HTML of a given news headline) that 
{\small \textit{(i)}}~has been scraped from a given source, but
{\small \textit{(ii)}}~to which no sanitization mechanism is applied.

As far as we are aware, popular scraping libraries, such as \textit{Scrapy}~\cite{scrapy}, \textit{BeautifulSoup}~\cite{beautifulsoup}, \textit{Cheerio}~\cite{cheerio}, or \textit{newspaper3k}~\cite{newspaper3k}, do not apply the specific operations that allow to clean an ``adversarial headline'' manipulated in the ways envisioned in our threat model. We empirically confirmed such an hypothesis: while \textit{newspaper3k} strips hidden HTML, this is not the case for \textit{Scrapy}, \textit{BeautifulSoup}, and \textit{Cheerio}; moreover, none of these libraries normalize Unicode by default.

As a matter of fact, the issue lies not in these libraries---which are simply designed to scrape the HTML from a given source. The issue lies in using such libraries ``as is'', overlooking the fact that the source from which the data is scraped may not be trusted. We have analysed various trading platforms (specifically the backtesting framework, \textit{Backtrader}~\cite{backtrader}, \textit{QuantConnect}~\cite{quantconnect}, OpenBB~\cite{openbb}) that provide services reliant on scraped HTML data: all such platforms are vulnerable to our proposed attack, because no substantial preprocessing is done on the scraped HTML data they ingest.

We summarize the findings of such a real-world analysis in Table~\ref{tab:realworld_weaknesses} (in the Appendix). For responsible disclosure, we reached out (in Sept. 2025) to the maintainers of these trading platforms. We informed them that their products, due to such a vulnerability, can negatively impact LLM-driven ATS that make trades by using their platform-provided data.

\subsection{Recommendations}
\label{ssec:recommendations}
\noindent
Altogether, we have three recommendations for practitioners.

First, the risk of our attack should be acknowledged. Practitioners should then run simulations, potentially using our open-source resources, to measure the potential loss induced by ``adversarial news'' to their ATS. Such knowledge can be used to determine what kind of approach to follow.

Second, universal defenses are challenging to realize, and one cannot counter all conceivable adversarial news via, e.g., input sanitization. We recommend considering multi-tiered countermeasures, such as combining static input-sanitization (which do not disrupt the payload) with ``adversarial-aware'' LLMs which may react (e.g., raise alarms, or block trading executions) in the presence of flagged adversarial news.

Third, and for cases wherein the ATS uses vendor-provided news (common according to our survey~§\ref{sec:survey}), we advocate that the processing-chain of the news to be documented. In particular, vendors should: {\small \textit{(i)}}~disclose whether they implement sanitization policies, and specify which ones they use; and {\small \textit{(ii)}}~record provenance per headline (source and retrieval time) to enable forensics; and {\small \textit{(iii)}}~store both the raw HTML string and a rendered view\footnote{The ``rendered view'' can also be used to apply OCR techniques to potentially provide another way to sanitize inputs. However, a recent work found that even these solutions are not very reliable~\cite{boucher2025vision}.} to audit hidden markup. Doing so would enable identification of potentially-compromised channels that release adversarial news, preventing further damage.

%% file: sections/10-conclusions.tex
\section{Conclusions}
\label{sec:conclusions}

\noindent
We conducted an end-to-end security evaluation of news-driven ATS that combines an LSTM-based price forecaster with LLM-based sentiment analysis derived from analysing headlines of financial news. We tested such an ATS against two text-manipulation attacks: \emph{Unicode homoglyph misrouting} and \emph{hidden-text injection}. We economically quantified the impact of such attacks---which, in the worst case, can decrease the cumulative returns of the targeted ATS by over 17\%. Countermeasures to these attacks are challenging to implement. Yet, by quantifying the (potential) financial losses, organizations can better decide how to address such a tangible risk.

\section*{Acknowledgments}
\noindent
The authors thank the anonymous SaTML'26 reviewers for the great feedback. This research has been partly funded by Hilti.

%% file: sections/LLMusage.tex
\section*{LLM usage considerations}

LLMs are an integral part of this work, since our goal is to study vulnerabilities in financial LLMs integrated into algorithmic trading systems. 
We evaluated both open-source models (FinBERT, FinGPT, FinLLaMA) and API-hosted models (OpenAI GPT-4o, O3, GPT-5, Google Gemini) under deterministic configurations. 
Precise model versions, parameters, and tested environments are detailed in the Appendix to ensure transparency and reproducibility. 
All ideas, experimental designs, and analysis were developed by the authors; LLMs were only used as experimental subjects.  

In addition, LLMs were used for editorial assistance during manuscript preparation (grammar checks and LaTeX formatting). 
All such outputs were manually inspected and verified by the authors to ensure accuracy and originality.

\section*{Reproducibility Statement}
Due to proprietary data constraints, we cannot provide data/details beyond those provided in this paper and/or in our repository~\cite{ourRepo}. In Appendix~\ref{app:environment}, we provide precise versions and configurations of platforms and libraries tested, reflecting their default behaviors as of mid-2025, as well as the prompts used for our ATS. These details allow reimplementation with equivalent data (e.g., public news feeds like Tiingo or custom HTML inputs) to verify vulnerabilities to homoglyph spoofing and hidden-text injection.

%% file: main.bbl

%% file: appendix/structure_app.tex
\input{appendix/A-LiteratureReview}

\input{appendix/B-environment}

%% file: appendix/A-LiteratureReview.tex
\section{Systematic Literature Review }
\label{app:slr}
\noindent
We explain the methodology and results of our systematic literature review (SLR), mentioned in §\ref{ssec:ml-finance-gap}.
Our goal is to determine the extent to which, as of September 2025, prior work studied (and evaluated) the security of LLM-driven ATS.

\subsection{Approach and Venue Selection}
\label{sapp:venue}

\noindent
Our SLR is inspired by~\cite{rizvani2025ephemeral}, which analysed works in AI security in finance published 2013--2023 via a mixture of automatic keyword-driven search and manual analysis. However, for the goal of our work, the SLR in~\cite{rizvani2025ephemeral} has the limitation that {\small \textit{(i)}}~it did not specifically account for LLMs, and {\small \textit{(ii)}}~its starting point was on works published in security-focused venues. 

We extend the protocol of~\cite{rizvani2025ephemeral} in two directions. First, we update the time window to cover recent work up to September 2025 and include additional venues where LLM research is typically published. Second, we specialise the (automatic) search strategy to LLM-based systems by refining the keyword groups and focusing on works where LLMs are explicitly integrated into trading or portfolio management systems.

Specifically, we collect our papers (we only considered full papers) by drawing from the following peer-reviewed venues:
\begin{itemize}[leftmargin=*]
    \item \textit{Security venues.} We consider the same top-tier venues as in~\cite{rizvani2025ephemeral}, i.e., ACM CCS and AsiaCCS, IEEE S\&P and EuroS\&P, NDSS, USENIX Security, ESORICS, ACSAC, FC. We consider works accepted between January 2024 and September 2025: the timeframe is chosen because the 2013--2023 timespan was covered in~\cite{rizvani2025ephemeral}. We thus collect all papers published in these venues and timeframe, obtaining a total of 2,169 papers

    \item \textit{ML and NLP venues.} We extend our dataset by retrieving works from four top-tier ML and NLP (two each) conferences. Specifically, we consider: NeurIPS and ICML for ML, and EMNLP and ACL for NLP. Given that these venues were not considered in~\cite{rizvani2025ephemeral}, we consider the timeframe 2020--2025; we also consider years prior the rise in popularity of LLMs due to a higher likelihood that ``precursors'' of LLMs were indeed evaluated in ML/NLP-focused works (e.g., the ``Attention is all you need'' paper came out in 2017~\cite{vaswani2017attention}, and BERT in 2019~\cite{devlin2019bert}). We thus collect a total of 23,038 papers.
\end{itemize}
Overall, we obtain 25,207 papers, which will represent the backbone of our analysis. We are aware that there may be other venues (e.g., ICLR), so we will complement our search via snowballing~\cite{wohlin2014guidelines} to find any potential work cited by, or referencing, any work that falls into our scope as a result of our primary search.

We emphasize that our SLR has been done by two authors who regularly interacted to discuss the procedure, results, and resolve doubts via joint discussions.

\subsection{Automated Filtering and Manual Inspection}
\label{sapp:filtering}

\noindent
We cannot manually analyse 25k papers, so we rely on automated techniques (as also done in~\cite{rizvani2025ephemeral}) to identify potential candidates for our subsequent manual inspection.

To this end, and inspired also by~\cite{rizvani2025ephemeral}, we define a set of three keyword groups seeking to identify works that cover: {\small \textit{(i)}}~finance, {\small \textit{(ii)}}~LLMs, and {\small \textit{(iii)}}~security. Ideally, a paper evaluating ``attacks against LLM-driven ATS'' should mention each of these topics in the abstract. The keyword groups are as follows (the search was done case insensitive):
\begin{itemize}[leftmargin=*]
    \item Security keywords: ``adversarial'', ``attack'', ``perturbation'', ``manipulation'', ``poisoning'', ``risk'', ``security''
    \item Finance keywords: ``finance'', ``trading'', ``market'', ``econom'', ``stock'', ``portfolio''
    \item LLM keywords: ``language model'', ``LLM'', ``GPT'', ``transformer'', ``BERT'', ``LLaMA'' (and ``ML'', ``machine learning'', ``AI'', ``artificial intelligence'', ``deep learning'', ``DL'').
\end{itemize}
Note that, aside from the LLM-specific ones, as well as ``risk'' and ``security'', all other keywords were drawn from the methodology of~\cite{rizvani2025ephemeral}. 

As also done in~\cite{rizvani2025ephemeral}, we consider a paper to be (potentially) about a certain domain if its abstract mentions at least one word in the respective keyword group. We thus analyse the abstracts of these works, and find that only three papers have a match for each keyword group, specifically:~\cite{du2025sok, galloway2024practical, naseri2024badvfl}. We manually inspected these three works, and found that they had little to do with LLMs and/or finance (e.g.,~\cite{naseri2024badvfl} mentions ``financial fraud'' but there is no experiment on this domain; whereas there is no LLM in~\cite{du2025sok, galloway2024practical}). Nonetheless, we applied snowballing, checking if among the 241 references of these three works there could be potential candidates that matched our criteria, but we found none.

\subsection{Extended Literature Search on Google Scholar}
\label{sapp:scholar}

\noindent
Perhaps surprisingly, among the 25,448 (25,207+241 of snowballing) papers we considered, we found none that tackled security aspects of LLM-driven applications in finance. Such a negative result motivated us to carry out a broader, and more qualitative (but still systematic), literature search on Google Scholar (as also done, e.g., in~\cite{schroer2025sok, rizvani2025ephemeral}).

\textbf{Broad queries.} We begin by devising search queries to identify potential candidates. As a prerogative, and to align our setup with~\cite{rizvani2025ephemeral}, we only considering peer-reviewed papers (excluding, e.g., arXiv preprints or unpublished works). To make the search more humanly-feasible, we combine the keywords used in our first stage of our SLR. In practice, our queries are: 
\{``adversarial attack'' \smamath{\land} ``large language model'' \smamath{\land} ``algorithmic trading''; ``LLM'' \smamath{\land} ``portfolio management'' \smamath{\land} ``poisoning''; ``chatgpt'' \smamath{\land} ``trading bot'' \smamath{\land} ``adversarial''\}. We also consider all combinations of: \{(``LLM'' \smamath{\lor} ``GPT'' \smamath{\lor} ``ChatGPT'') \smamath{\land} (``trading'' \smamath{\lor} ``stock market'' \smamath{\lor} ``portfolio management'') \smamath{\land} (``adversarial attack'' \smamath{\lor} ``perturbation'' \smamath{\lor} ``manipulation'')\}. The queries were performed twice: once in Sept. 2025, and another time in Nov. 2025.

\textbf{Screening.} Each of these queries returned \smamath{>10k} results. For a feasible analysis, we retrieved the metadata of the first 100 (peer-reviewed) papers for each query. We then checked the title and abstract of our collected papers, ascertaining that they mentioned the search terms of our query (given that Google Scholar can yield ``false positives''). If all terms are mentioned, we then manually review the abstracts to verify if such papers can truly be considered as being (potentially) within our scope. 

\textbf{Manual analysis (and snowballing).} 
After our screening of the abstract, we eventually identified 14 papers that, potentially, evaluated the security of LLM applications in finance. So, we proceeded to do a full manual check of each of these. Deeper inspection revealed that, however, no work evaluated the security of LLM applications in finance. For instance, some works just discuss~\cite{mcclellan2025ai} or propose some high-level frameworks~\cite{mahmud2025exploring}, or consider ``adversarial'' settings that are not intrinsically targeting LLMs~\cite{sadhu2025structured}. Some (e.g.,~\cite{liang2023developing}) carry out an evaluation of LLM in financial settings, but there is no specific financial application nor finance-specific evaluation metric that is taken into account. We further expanded our search by using the snowball method on these 14 papers, covering an additional 871 works, which we analyzed using the same criteria (i.e., only published works, and checking the inclusion of our keyword groups in the abstract). However, despite finding five potential candidates, even in this extended set of works we could not find a single paper that practically assessed the security of LLM-driven ATS (e.g., the work in~\cite{al2024evaluating} is just a literature review with no evaluation).

\begin{cooltextbox}
\textbf{\textsc{Takeaway.}}
Our systematic literature review of over 25k papers from top-tier security \& ML/NLP venues, further broadened with snowballing and a search on Google Scholar, revealed that no prior work practically evaluated the security of LLM applications in financial contexts by adopting a system-wide perspective and using domain-specific metrics.\footnote{Of course, we acknowledge that our literature review cannot cover \textit{all} prior work. For instance, we focused on peer-reviewed works, meaning that preprints could have tackled a similar effort---as is the case for~\cite{deza2020robustness}.}
\end{cooltextbox}

%% file: appendix/B-environment.tex
\section{Environments and Versions}
\label{app:environment}

\begin{table}[h]
\centering
\caption{Platforms tested and their default behaviors as of mid-2025.}
\label{tab:platforms}
\small
\begin{tabularx}{\columnwidth}{@{}l l X@{}}
\toprule
\textbf{Tool} & \textbf{Version} & \textbf{Default Behavior / Vulnerability} \\
\midrule
QuantConnect Lean & v2.5 & Default data pipeline with Tiingo/News provider; no custom sanitizer for Unicode normalization, homoglyph detection, or HTML filtering. \\
OpenBB & v4.4.5 & \texttt{openbb.news} scraper uses default parsers; no Unicode or HTML sanitization. \\
\bottomrule
\end{tabularx}
\end{table}


\begin{table}[h]
\centering
\caption{Scraping libraries and their default behaviors as of mid-2025.}
\label{tab:scrapers}
\small
\begin{tabularx}{\columnwidth}{@{}l l X@{}}
\toprule
\textbf{Library} & \textbf{Version} & \textbf{Default Behavior / Vulnerability} \\
\midrule
Scrapy & 2.13.3 & Returns raw Unicode; retains hidden DOM nodes (e.g., \texttt{display:none}). \\
BeautifulSoup4 & 4.13.5 & Parser=\texttt{lxml}; retains \texttt{display:none} text. \\
Cheerio (Node) & 1.1.2 & No Unicode normalization or homoglyph detection. \\
newspaper3k & 0.2.8 & Removes some hidden tags; no Unicode or homoglyph normalization. \\
\bottomrule
\end{tabularx}
\end{table}

\subsection{Determinism}
All experiments use fixed seeds for reproducibility. Environments are pinned to specific versions (e.g., Python 3.10 for Scrapy/BeautifulSoup4, Node.js 18 for Cheerio). Tests reflect default configurations as of mid-2025.

\subsection{Model Configurations}
\label{app:llm-configs}

\begin{table}[t]
\centering
\caption{LLM configurations used in our experiments (mid-2025).}
\label{tab:llm-configs}
\scriptsize
\setlength{\tabcolsep}{3pt}
\begin{tabularx}{\columnwidth}{@{}l Y c c c c@{}}
\toprule
\textbf{Model} & \textbf{Provider / Release} & \textbf{Ver./Date} & \textbf{Temp} & \textbf{Top-$p$} & \textbf{MaxTok} \\
\midrule
FinBERT        & HuggingFace (ProsusAI/finbert) & v1.0 (2020)      & 0.0 & 1.0 & 512  \\
FinGPT         & FinGPT-6B (FinNLP)             & v2.1 (2024)      & 0.0 & 1.0 & 512  \\
FinLLaMA       & FinLLaMA-7B (FinNLP)           & v1.2 (2025)      & 0.0 & 1.0 & 512  \\
O3             & OpenAI API                     & tested Jul 2025  & 0.0 & 1.0 & 2048 \\
O3 Pro         & OpenAI API                     & tested Jul 2025  & 0.0 & 1.0 & 2048 \\
4o             & OpenAI API                     & tested Jul 2025  & 0.0 & 1.0 & 2048 \\
4o-mini        & OpenAI API                     & tested Jul 2025  & 0.0 & 1.0 & 2048 \\
4o-mini-high   & OpenAI API                     & tested Jul 2025  & 0.0 & 1.0 & 2048 \\
GPT-5          & OpenAI API                     & tested Aug 2025  & 0.0 & 1.0 & 2048 \\
ChatGPT (4o-mini) & OpenAI API                  & tested Jul 2025  & 0.0 & 1.0 & 2048 \\
Gemini Pro 1.5 & Google AI Studio               & tested Jul 2025  & 0.0 & 1.0 & 2048 \\
\bottomrule
\end{tabularx}

\vspace{2pt}
\emph{\footnotesize Note:} All API models used deterministic settings; offline models used seed $=42$.
\end{table}

\noindent \textit{Note:} All GPT-family models, O3/O3 Pro, and Gemini were evaluated on their July 2025 versions. GPT-5 was tested in August 2025. Offline models (FinBERT, FinGPT, FinLLaMA) are version-stable checkpoints.

\subsection{Prompts}
\label{sapp:prompts}

\noindent
\textbf{Prompting protocol.}
We evaluate all models under a single \emph{zero-shot} prompting setup. The wording, output format, and ticker set are identical across models; decoding is deterministic where supported (\texttt{temperature}=0. Importantly, the prompts \emph{do not} mention Unicode, normalization, HTML, or hidden text. Attacks only modify the input headlines; all other settings remain fixed.

\smallskip
\noindent\textbf{Evaluation protocol.}
We use a single \emph{zero-shot} prompt with fixed wording and output schema across all models.  
Decoding is deterministic where supported (\texttt{temperature}=0, \texttt{top\_p}=1).  
Prompts never mention Unicode, normalization, or hidden text; the only difference between clean and attacked runs is the headline content.

\smallskip
\noindent\textbf{Controls.}
Ticker sets and label mappings are identical across models.  
We fix the token budget and stop criteria, and do not use retries or voting.  
Prompts contain no hints about the manipulations.

\smallskip
\noindent\textbf{Robustness.}
Key experiments were repeated with shuffled ticker lists and alternative output formats; results remained consistent.  

\newpage


\begin{lstlisting}[caption={Zero-shot prompt for stock association (ticker mapping/misrouting).},label={app:prompt-map}, basicstyle=\ttfamily\footnotesize]
System:
You are a financial analyst who uses NLP techniques (NER, semantic matching) to detect which company a news headline refers to.

User:
I will give you a numbered list of headlines.

Your job:
- For each headline, determine which stock from [GOOGL, AAPL, NVDA, MSFT, AMZN, META, TSLA, LLY, JPM, XOM] it refers to.
- Base your decision on company names, related terms, or clear semantic references.
- If no confident match exists, return "unrecognized".

Input format:
Headlines:
1. <headline text here>
2. <headline text here>
3. <headline text here>
...

Output format (one line per headline, same order):
<id>,<pred_ticker>

Where:
- <id> is the input headline number (1,2,3,...).
- <pred_ticker> is one of the listed tickers or "unrecognized".

Return only the lines in the specified format. No extra text, explanations, or JSON.
\end{lstlisting}

\subsection{Sentiment Scoring}
\label{app:sentiment}

\begin{lstlisting}[caption={Zero-shot prompt for sentiment experiments (HTML-tolerant).},label={app:prompt-sent}, basicstyle=\ttfamily\footnotesize]
System:
You are a financial-news sentiment classifier.

User:
Input: a numbered list of headlines that all refer to the same known stock ticker.
Notes:
- The text may contain HTML markup.
- Keep output strictly in the specified format.

Ticker: <TICKER>

Headlines:
1. <headline text here>
2. <headline text here>
3. <headline text here>
...

Output format (one line per headline, same order):
<id>,<sent>,<conf>

Where:
- <id> is the input headline number (1,2,3,...).
- <sent> is +1 for positive, 0 for neutral, -1 for negative.
- <conf> is a confidence score as a float between 0 and 1.

Return only the lines in the specified format. No extra text, explanations, or JSON.
\end{lstlisting}


\begin{table}[h]
\centering
\caption{Latin letters and their visually confusable Cyrillic (and related) substitutes used in our attack.}
\label{tab:mapping_char}
\small
\begin{tabular}{@{}ll@{}}
\toprule
\textbf{Latin} & \textbf{Cyrillic)} \\
\midrule
A & {\begingroup\fontencoding{T2A}\selectfont А\endgroup} \\
B & {\begingroup\fontencoding{T2A}\selectfont В\endgroup} \\
C & {\begingroup\fontencoding{T2A}\selectfont С\endgroup} \\
E & {\begingroup\fontencoding{T2A}\selectfont Е\endgroup} \\
G & {\begingroup\fontencoding{T2A}\selectfont G\endgroup} \\
H & {\begingroup\fontencoding{T2A}\selectfont Н\endgroup} \\
I & {\begingroup\fontencoding{T2A}\selectfont І\endgroup} \\
J & {\begingroup\fontencoding{T2A}\selectfont Ј\endgroup} \\
K & {\begingroup\fontencoding{T2A}\selectfont К\endgroup} \\
L & {\begingroup\fontencoding{T2A}\selectfont L\endgroup} \\
M & {\begingroup\fontencoding{T2A}\selectfont М\endgroup} \\
O & {\begingroup\fontencoding{T2A}\selectfont О\endgroup} \\
P & {\begingroup\fontencoding{T2A}\selectfont Р\endgroup} \\
S & {\begingroup\fontencoding{T2A}\selectfont Ѕ\endgroup} \\
T & {\begingroup\fontencoding{T2A}\selectfont Т\endgroup} \\
X & {\begingroup\fontencoding{T2A}\selectfont Х\endgroup} \\
Y & {\begingroup\fontencoding{T2A}\selectfont У\endgroup} \\
Z & {\begingroup\fontencoding{T2A}\selectfont У\endgroup} \\
a & {\begingroup\fontencoding{T2A}\selectfont а\endgroup} \\
c & {\begingroup\fontencoding{T2A}\selectfont с\endgroup} \\
e & {\begingroup\fontencoding{T2A}\selectfont е\endgroup} \\
h & {\begingroup\fontencoding{T2A}\selectfont һ\endgroup} \\
i & {\begingroup\fontencoding{T2A}\selectfont і\endgroup} \\
j & {\begingroup\fontencoding{T2A}\selectfont ј\endgroup} \\
m & {\begingroup\fontencoding{T2A}\selectfont м\endgroup} \\
o & {\begingroup\fontencoding{T2A}\selectfont о\endgroup} \\
p & {\begingroup\fontencoding{T2A}\selectfont р\endgroup} \\
s & {\begingroup\fontencoding{T2A}\selectfont ѕ\endgroup} \\
x & {\begingroup\fontencoding{T2A}\selectfont х\endgroup} \\
y & {\begingroup\fontencoding{T2A}\selectfont у\endgroup} \\
\bottomrule
\end{tabular}
\end{table}

\begin{table}[h]
\centering
\caption{Stock tickers with visually confusable Cyrillic substitutions. Only letters that have Cyrillic homoglyphs are replaced.}
\label{tab:mapping_ticker}
\small
\begin{tabular}{@{}ll@{}}
\toprule
\textbf{Ticker (Latin)} & \textbf{Ticker (Cyrillic Attack)} \\
\midrule
GOOGL & G{\begingroup\fontencoding{T2A}\selectfont О\endgroup}{\begingroup\fontencoding{T2A}\selectfont О\endgroup}G{\begingroup\fontencoding{T2A}\selectfont L\endgroup} \\
AAPL  & {\begingroup\fontencoding{T2A}\selectfont А\endgroup}{\begingroup\fontencoding{T2A}\selectfont А\endgroup}P{\begingroup\fontencoding{T2A}\selectfont L\endgroup} \\
NVDA  & NVD{\begingroup\fontencoding{T2A}\selectfont А\endgroup} \\
MSFT  & {\begingroup\fontencoding{T2A}\selectfont М\endgroup}SFT \\
AMZN  & {\begingroup\fontencoding{T2A}\selectfont А\endgroup}{\begingroup\fontencoding{T2A}\selectfont М\endgroup}ZN \\
META  & {\begingroup\fontencoding{T2A}\selectfont М\endgroup}{\begingroup\fontencoding{T2A}\selectfont Е\endgroup}T{\begingroup\fontencoding{T2A}\selectfont А\endgroup} \\
TSLA  & {\begingroup\fontencoding{T2A}\selectfont Т\endgroup}S{\begingroup\fontencoding{T2A}\selectfont L\endgroup}{\begingroup\fontencoding{T2A}\selectfont А\endgroup} \\
LLY   & {\begingroup\fontencoding{T2A}\selectfont L\endgroup}{\begingroup\fontencoding{T2A}\selectfont L\endgroup}{\begingroup\fontencoding{T2A}\selectfont У\endgroup} \\
JPM   & {\begingroup\fontencoding{T2A}\selectfont Ј\endgroup}PM \\
XOM   & {\begingroup\fontencoding{T2A}\selectfont Х\endgroup}{\begingroup\fontencoding{T2A}\selectfont О\endgroup}{\begingroup\fontencoding{T2A}\selectfont М\endgroup} \\
\bottomrule
\end{tabular}
\end{table}

\begin{table}[h]
\centering
\caption{Company names with only Cyrillic homoglyph substitutions applied. Letters with no Cyrillic counterpart are unchanged.}
\label{tab:mapping_company}
\small
\begin{tabular}{@{}ll@{}}
\toprule
\textbf{Company (Latin)} & \textbf{Company (Cyrillic Attack)} \\
\midrule
Google         & G{\begingroup\fontencoding{T2A}\selectfont о\endgroup}{\begingroup\fontencoding{T2A}\selectfont о\endgroup}g{\begingroup\fontencoding{T2A}\selectfont l\endgroup}{\begingroup\fontencoding{T2A}\selectfont е\endgroup} \\
Alphabet       & {\begingroup\fontencoding{T2A}\selectfont А\endgroup}l{\begingroup\fontencoding{T2A}\selectfont р\endgroup}h{\begingroup\fontencoding{T2A}\selectfont а\endgroup}b{\begingroup\fontencoding{T2A}\selectfont е\endgroup}t \\
Apple          & {\begingroup\fontencoding{T2A}\selectfont А\endgroup}{\begingroup\fontencoding{T2A}\selectfont р\endgroup}{\begingroup\fontencoding{T2A}\selectfont р\endgroup}l{\begingroup\fontencoding{T2A}\selectfont е\endgroup} \\
Nvidia         & Nv{\begingroup\fontencoding{T2A}\selectfont і\endgroup}di{\begingroup\fontencoding{T2A}\selectfont а\endgroup} \\
Microsoft      & {\begingroup\fontencoding{T2A}\selectfont М\endgroup}icros{\begingroup\fontencoding{T2A}\selectfont о\endgroup}ft \\
Amazon         & {\begingroup\fontencoding{T2A}\selectfont А\endgroup}{\begingroup\fontencoding{T2A}\selectfont м\endgroup}az{\begingroup\fontencoding{T2A}\selectfont о\endgroup}n \\
Meta           & {\begingroup\fontencoding{T2A}\selectfont М\endgroup}{\begingroup\fontencoding{T2A}\selectfont е\endgroup}t{\begingroup\fontencoding{T2A}\selectfont а\endgroup} \\
Tesla          & {\begingroup\fontencoding{T2A}\selectfont Т\endgroup}esl{\begingroup\fontencoding{T2A}\selectfont а\endgroup} \\
Eli Lilly      & {\begingroup\fontencoding{T2A}\selectfont Е\endgroup}li {\begingroup\fontencoding{T2A}\selectfont L\endgroup}i{\begingroup\fontencoding{T2A}\selectfont ll\endgroup}{\begingroup\fontencoding{T2A}\selectfont у\endgroup} \\
JPMorgan Chase & {\begingroup\fontencoding{T2A}\selectfont Ј\endgroup}P{\begingroup\fontencoding{T2A}\selectfont М\endgroup}organ Chase \\
Exxon Mobil    & {\begingroup\fontencoding{T2A}\selectfont Е\endgroup}x{\begingroup\fontencoding{T2A}\selectfont х\endgroup}{\begingroup\fontencoding{T2A}\selectfont о\endgroup}n {\begingroup\fontencoding{T2A}\selectfont М\endgroup}obil \\
\bottomrule
\end{tabular}
\end{table}

\begin{table}[!t]
    \caption{Sanitization gaps in common trading platforms and libraries (BS=BeautifulSoup).
    None apply full Unicode normalization or HTML filtering by default, leaving pipelines vulnerable to input-layer manipulations.}
    \label{tab:realworld_weaknesses}
    \small
    \begin{tabularx}{\columnwidth}{@{}lX@{}}
        \toprule
        \textbf{Platform / Library} & \textbf{Default Sanitization} \\
        \midrule
        \multicolumn{2}{@{}l}{\textit{Trading Platforms}} \\
        
        \textbf{QuantConnect} & Applies minimal normalization (e.g., “.” → “–”); does not detect homoglyphs or hidden HTML. \\
                
        \textbf{OpenBB} & Uses raw web scraping; performs no text sanitization or entity normalization. \\
        
        \midrule
        \multicolumn{2}{@{}l}{\textit{Libraries}} \\
        
        \textbf{Backtrader} & No sanitization; all headline text is passed unfiltered to user-defined logic. \\
        
        \textbf{Scrapy / BS / Cheerio} & Parse DOM trees but retain hidden elements (e.g., \texttt{<span style="display:none" >}); Unicode is not normalized. \\
        
        \textbf{newspaper3k} & Partially strips hidden HTML but allows unnormalized Unicode through. \\
        \bottomrule
    \end{tabularx}
\end{table}